\documentclass[aps,prd,notitlepage,reprint,twocolumn,showpacs,superscriptaddress,nofootinbib,floatfix]{revtex4-1}

\usepackage{amsmath,amssymb,graphicx}
\usepackage{mathrsfs}
\usepackage{bbm}
\usepackage{bbold}
\usepackage{color}
\usepackage{hyperref}

\newcommand{\be}{\begin{equation}}
\newcommand{\ee}{\end{equation}}
\newcommand{\bi}{\begin{itemize}}
\newcommand{\ei}{\end{itemize}}
\newcommand{\bea}{\begin{eqnarray}}
\newcommand{\eea}{\end{eqnarray}}

\definecolor{bc}{rgb}{0, 0.7, 0.0}

\newcommand{\ud}{\mathrm{d}}
\newcommand{\LCp}{{\scriptscriptstyle +}}
\newcommand{\LCpm}{{\scriptscriptstyle \pm}}

\newcommand{\LCperp}{{\scriptscriptstyle \perp}}

\usepackage[T1]{fontenc} \usepackage[latin1]{inputenc}

\begin{document} 

\title{Classical and quantum particle dynamics in univariate background fields}

\author{T.~Heinzl}
\email{thomas.heinzl@plymouth.ac.uk}
\affiliation{School of Computing and Mathematics, University of Plymouth, PL4 8AA, UK}

\author{A.~Ilderton}
\email{anton.ilderton@chalmers.se}
\affiliation{Department of Physics, Chalmers University, 41296 Gothenburg, Sweden}

\author{B.~King}
\email{b.king@plymouth.ac.uk}
\affiliation{School of Computing and Mathematics, University of Plymouth, PL4 8AA, UK}

\begin{abstract}
We investigate deviations from the plane wave model in the interaction of charged particles with strong electromagnetic fields. A general result is that integrability of the dynamics is lost when going from lightlike to timelike or spacelike field dependence. For a special scenario in the classical regime we show how the radiation spectrum in the spacelike (undulator) case becomes well-approximated by the plane wave model in the high energy limit, despite the two systems being Lorentz inequivalent. In the quantum problem, there is no analogue of the WKB-exact Volkov solution. Nevertheless, WKB and uniform-WKB approaches give good approximations in all cases  considered. Other approaches that reduce the underlying differential equations from second to first order are found to miss the correct physics for situations corresponding to barrier transmission and wide-angle scattering.
\end{abstract}

\maketitle

\section{Introduction}\label{intro}
Strong fields provide new opportunities for measuring unobserved processes both within the Standard Model and beyond. Scattering amplitude calculations for these processes require dressed particle wavefunctions in order to  describe the interaction with the strong (background) field. Due to the high field strengths involved, however, the wavefunctions cannot be constructed using the standard method of perturbation in the coupling. Thus one must either use nonperturbative approximations, or work with special cases where exact solutions exist.

The latter situation is typified by intense laser-matter interactions (for introductions and reviews see~\cite{Ritus,Marklund:2006my,Heinzl:2008an,Gies:2008wv,Dunne:2008kc,DiPiazza:2011tq,King:2015tba}), where the plane wave model of a laser field allows for a fully analytical treatment~\cite{Volkov,Reiss:1962,Nikishov:1963,Nikishov:1964a,Narozhnyi:1964,Goldman:1964,Kibble:1965zza}. The model is powerful but restrictive, being unable to account for effects due to the spatial geometry of a realistic laser pulse. This limits our ability to predict and analyse experimental results -- indeed the primary source of phenomenological results for the complex interactions of focussed laser fields with electron bunches and solid targets are PIC-simulations, reviewed in~\cite{Gonoskov:2014mda}.

In order to gain analytical insights, confirm numerical results, and improve experimental predictions, we must go beyond the plane wave model~\cite{Mendonca,Raicher:2013cja,DiPiazza:2013vra,Oertel:2015yma}. This task is as challenging 
as it is open-ended. We therefore approach it here by restricting ourselves to a 
specific class of extensions beyond the plane wave model, by retaining the 
single-variable ($k_\mu x^\mu$) field dependence, but generalising from lightlike $k_\mu$ to 
arbitrary~$k_\mu$.

There are only three Lorentz-inequivalent cases. For $k^2=0$ we have plane waves, and both the 
quantum and classical dynamics are integrable in the sense that the classical equations of motion, 
as well as the Dirac and Klein-Gordon equations, are exactly solvable due to the presence of 
sufficiently many conserved quantities. $k^2<0$ may describe an undulator~\cite{Jackson:1998nia} or 
a plane wave in a medium with refractive index $n_r>~1$~\cite{Cronstrom:1977}, depending on the 
chosen~$k_\mu$ or, equivalently, the chosen frame: these two systems are boost-equivalent. 
Similarly, $k^2>0$ can describe a plane wave in a medium of refractive index 
$n_r<1$~\cite{Cronstrom:1977,Becker:1977},  or a time-dependent electric field, depending on the 
chosen frame~\cite{Bulanov:2003aj}. The latter case can also be obtained by restricting to a 
magnetic field node of a standing wave depending on two lightfront variables.

For $k^2\not=0$ there are no general solutions to the Klein-Gordon or Dirac equations. When 
building approximate solutions there are broadly two approaches to consider. Either one can look for 
general approximations, which hold for all field shapes as the Volkov solution for plane waves 
does~\cite{Volkov}, or one can look for approximations specific to a particular field shape. In the 
case of $k^2>0$ both the former~\cite{Mendonca} and 
latter~\cite{Raicher:2013cja,Raicher:2015ara,Raicher:2016bbx} have recently begun to attract 
attention. Here we will mainly focus on the case $k^2<0$. The approaches we describe can also be 
applied to $k^2>0$, although the physics is different~\cite{Becker:1977}. For an analysis of the 
Klein-Gordon equation in this case, in the context of standing waves, see \cite{Hu:2015}.

We will consider several different approaches to the problem of building accurate approximations, examine the connections between them, the situations in which they are applicable, and their sensitivity to kinematic parameters and field shape.

This paper is organised as follows. We begin in Sect.~\ref{SECT:KLASS} with a classical comparison of particle dynamics in fields with $k^2=0$ and $k^2\not=0$. We compare the emission spectra for electrons in a laser and in an undulator, and explain why the spectra are similar in the high-energy limit, despite the two cases being Lorentz-inequivalent. In Sect.~\ref{SECT:FIRST} we turn to the quantum problem, which is to solve the Klein Gordon equation in the chosen background field (we consider only scalar QED, for simplicity). We review existing approaches and present an approach based on `reduction of order' as applied by Landau and Lifshitz to radiation reaction. In Sect.~\ref{SECT:SECOND} we consider a different approach: by rewriting the Klein-Gordon equation as a Schr\"odinger equation we are able to use intuition from quantum mechanics to develop accurate approximate wavefunctions. Examples for incoming and outgoing scattering states are given. We use the intuition built up from this in Sect.~\ref{SECT:COMP} to re-analyse the literature and reduction of order approaches. We conclude in Sect.~\ref{SECT:CONCS}.

\section{Classical dynamics: lightlike vs. spacelike field dependence}\label{SECT:KLASS}
The field strength of our background is
\be\label{FIELD}
	eF_{\mu\nu}(x) = \big(k_\mu l^j_\nu - l^j_\mu k_\nu\big) f'_j(k.x) \;,
\ee
where $k.l^j=0$, $l_i.l_j = -\delta_{ij}$ and the $f'_j$ define the shape and amplitude of the 
field. For $k^2=0$ (\ref{FIELD}) describes a plane wave (in all frames). For $k^2<0$, we can go to a 
frame where $k_\mu=\omega_u(0,0,0,1)_{\mu}$, in which case (\ref{FIELD}) describes a static but 
position-dependent magnetic field as may be found in an undulator. If we then boost along~$z$,  
$k_\mu \to \omega_u\gamma\beta(1,0,0,1/\beta)_{\mu}$ and in this frame (\ref{FIELD}) describes a 
plane wave propagating in a medium with refractive index 
$n_r=1/\beta>1$~\cite{Cronstrom:1977,Becker:1977}. Similarly, for $k^2>0$ we can take 
$k_\mu=\omega_u(1,0,0,0)_{\mu}$, which describes a time-dependent but homogeneous electric field, 
and boosting gives a plane wave in a medium with $n_r=\beta <1$. 

As $\gamma\to\infty$ the frequencies in the boosted frames increase and $k_\mu/k^0$ approaches 
$(1,0,0,1)_{\mu}$, i.e.~the refractive index approaches unity. However, the plane wave case cannot 
be recovered by boosting, as $k^2$ is invariant. This is made explicit by introducing the 
boost rapidity $\zeta = \cosh^{-1}\gamma$, which enables us to write, 
for $k^2<0$,
\be\label{kprime}
	k \to \frac{1}{2}\omega_u e^\zeta (1,0,0,1) - \frac{1}{2}\omega_u e^{-\zeta} (1,0,0,-1) \;.
\ee
The `lightlike limit' $\zeta\to\infty$ would correspond to dropping the second term in (\ref{kprime}), whereupon ${k}^2$ would vanish. As this cannot be achieved with a boost, the emission spectra from e.g.~charges in undulators and charges in lasers cannot be equivalent~\cite{Harvey:2012ie}. However, the spectra do become similar at e.g.~high energy, or for large boosts, because even though $k^2$ is invariant, the contraction of $k_\mu$ with other vectors can clearly be dominated by the first term in (\ref{kprime}). Let us then compare the emission spectrum of an electron in fields with $k^2<0$ and $k^2=0$.

The spectral density of radiation with momentum $k'_\mu$ ($k'^2=0$)~is
\be
	\frac{\ud^3N}{\ud\Omega\ud\omega'} = -\frac{\omega'}{8\pi^3}  |j(k')|^2 \;,
\ee
where $j_\mu$ is the Fourier-transformed classical current. The notation $N$ refers to the fact 
that the spectral density becomes the average number of emitted photons in QED. In terms of proper 
time $\tau$, orbit $x^\mu$ and kinetic momentum~$\pi^\mu=m \dot{x}^\mu$ of the emitting particle, 
the current is
\be
	j_\mu(k') = \frac{e}{m}\int\!\ud\tau\; \pi_\mu(\tau) e^{ik'.x(\tau)} \;.
\ee
Hence the classical orbit is required. Writing $\phi\equiv k.x$ the 
Lorentz equation in the field (\ref{FIELD}) has the first integral
\be\label{p-gissning}
	\pi_\mu(\tau) = p_\mu - a_\mu(\phi(\tau)) + \frac{k.p}{k^2} k_\mu (s-1)\;,
\ee
in which the `work done' $a_{\mu}$ by the field is
\be\label{potential-val}
	a_\mu(\phi) := \int\limits_{\phi_0}^{\phi}\!\ud\varphi\;  f'_j(\varphi) l^j_\mu\;,
\ee
$p_\mu$ is the momentum at some chosen $\phi_0$ and
\be\label{L}\begin{split}
	s(\phi) &:=\sqrt{1 +\frac{2k^2}{k.p}u(\phi)} \;,\quad u(\phi) := \frac{ 2 p.a(\phi) - a^2(\phi)}{2k.p}  \;.
\end{split}
\ee
The sign of $s$ is determined by the condition that $\pi\to p$ as we turn the field off, and the 
`potential' $u(\phi)$ has the same form as in the plane wave case. Note that the momentum is at this 
stage only an implicit function of $\tau$: we also need to identify $\phi$ as a function of $\tau$ 
by solving the equation
\be\label{disc}
	\dot{\phi}(\tau)  = \frac{k.p}{m} s\big(\phi(\tau)\big) \;,
\ee
which amounts to performing part of the second integration to obtain the orbits. (Unlike in the plane wave case, $k.\pi$ is not conserved.) This can be done analytically only for special choices of $a_\mu(\phi)$ and kinematics (see below). The condition that $s\in\mathbb{R}$ restricts the range of positions $\phi$ accessible by the particle. The trajectory may be parameterised by $\phi$ \textit{only} if the discriminant in $s$ is positive for all $\phi$: $\phi(\tau)$ is then a monotonic function and we may trade $\tau$ for $\phi$. By taking $k^2\to0$ via e.g.~a coordinate rotation~\cite{Hornbostel:1991qj,Ji:2001xd,Ji:2012ux,Ilderton:2015qda} (not a boost), (\ref{p-gissning}), (\ref{L}) and (\ref{disc}) reduce to $k.\dot{x} = k.p/m \implies k.x = k.p\tau/m$ and
\be
	\pi_\mu(\tau) = p_\mu - a_\mu(k.x(\tau)) + u(k.x(\tau)) k_\mu \;,
\ee
which can straightforwardly be integrated again and yields the charge orbit motion in a plane wave~\cite{Landau:1982dva,Sarachik:1970ap}.

In order to make a connection with some of the examples below, in the quantum theory, consider the 
particular but familiar case of a head-on collision between an electron and a monochromatic, 
circularly polarised field\footnote{This case should be considered as the infinite duration limit of 
a long pulse, just as for the plane wave case, see~\cite{Kibble:1965zza}.}; in the frame where 
$k_\mu\sim(0,0,0,1)_{\mu}$, this is the field of a helical wiggler~\cite{Alferov:1979tp}. In this 
case $s$ is constant,
\be
	s \to \sqrt{1 + \frac{a_0^2m^2}{k.p^2}k^2} \;,
\ee
and the classical equation of motion is integrable, assuming parameters such that the discriminant 
is positive. The calculation of the spectral density is very similar to that 
in~\cite[\S101]{LL-QED}, so we skip to the final result. This is
\be\begin{split}\label{D3N}
	\frac{\ud^3N}{\ud\Omega\ud\omega'} = \frac{\alpha}{2\pi} \frac{m^2}{k.p^2s^2}V \sum\limits_{n>0} \omega' \delta\bigg(\frac{k'.q}{s k.p}-n\bigg) \mathcal{J}_n(z) \;,
\end{split}
\ee
where $z=a_0 m \omega'\sin\theta/|sk.p|$, $\theta$ is the emission angle relative to ${\bf k}$, the cycle-averaged (``quasi'') momentum~$q_\mu$~is
\be\label{q1}
	q_\mu = p_\mu + \frac{k.p}{k^2}(s-1)k_\mu \;,
\ee
and 
\be\label{J}
	\mathcal{J}_n = -2 J_n^2 + a_0^2 (J_{n+1}^2+J_{n-1}^2-2J_n^2) \;.
\ee
As in the plane wave case, periodicity results in a spectrum composed of discrete harmonics labelled by integer~$n$, and defined by the support of the delta function in (\ref{D3N}) which, we observe for later, depends on the quasi-momentum (\ref{q1}).  

\begin{figure}[t!]
\includegraphics[width=\columnwidth]{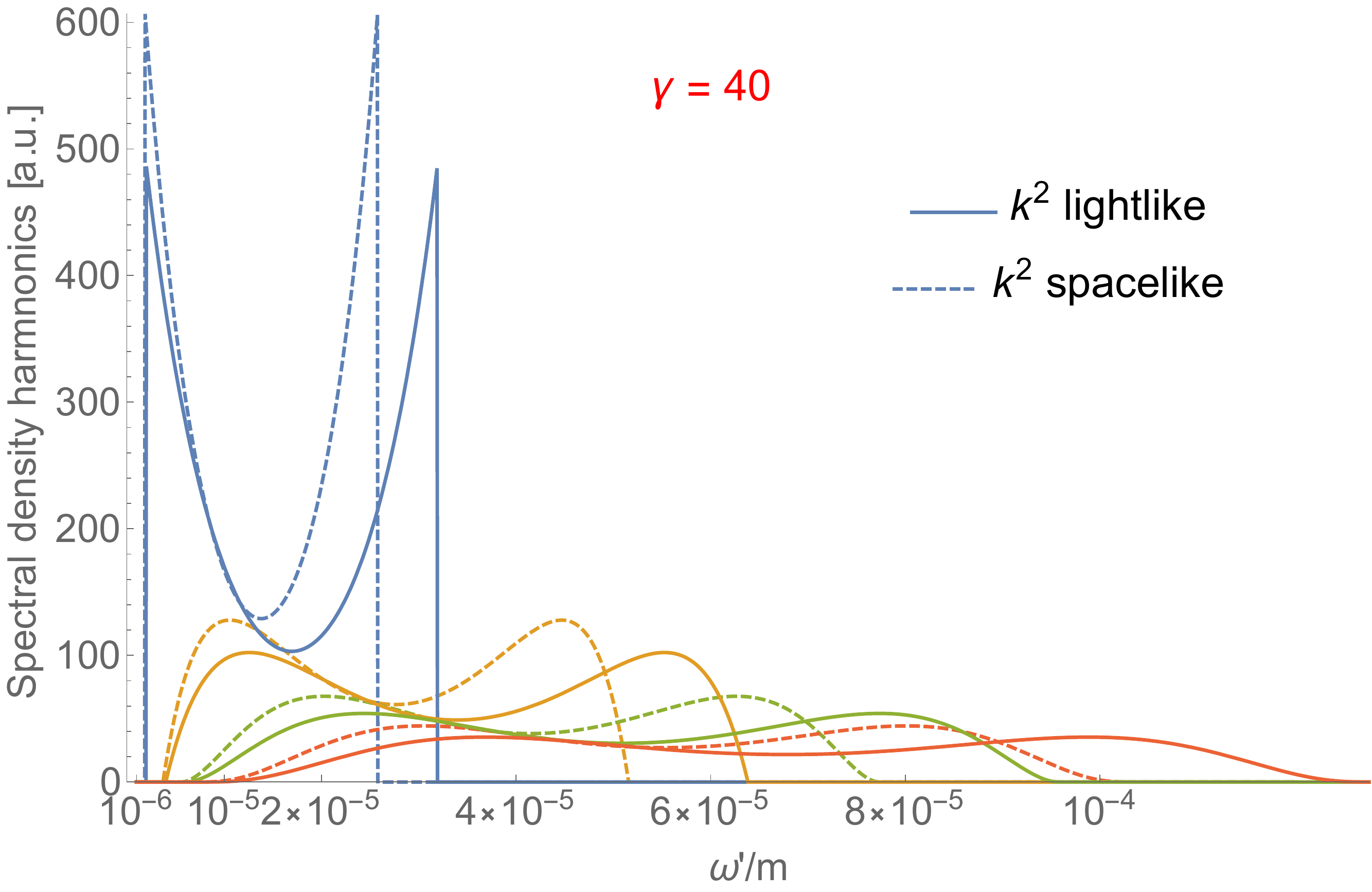}
\includegraphics[width=\columnwidth]{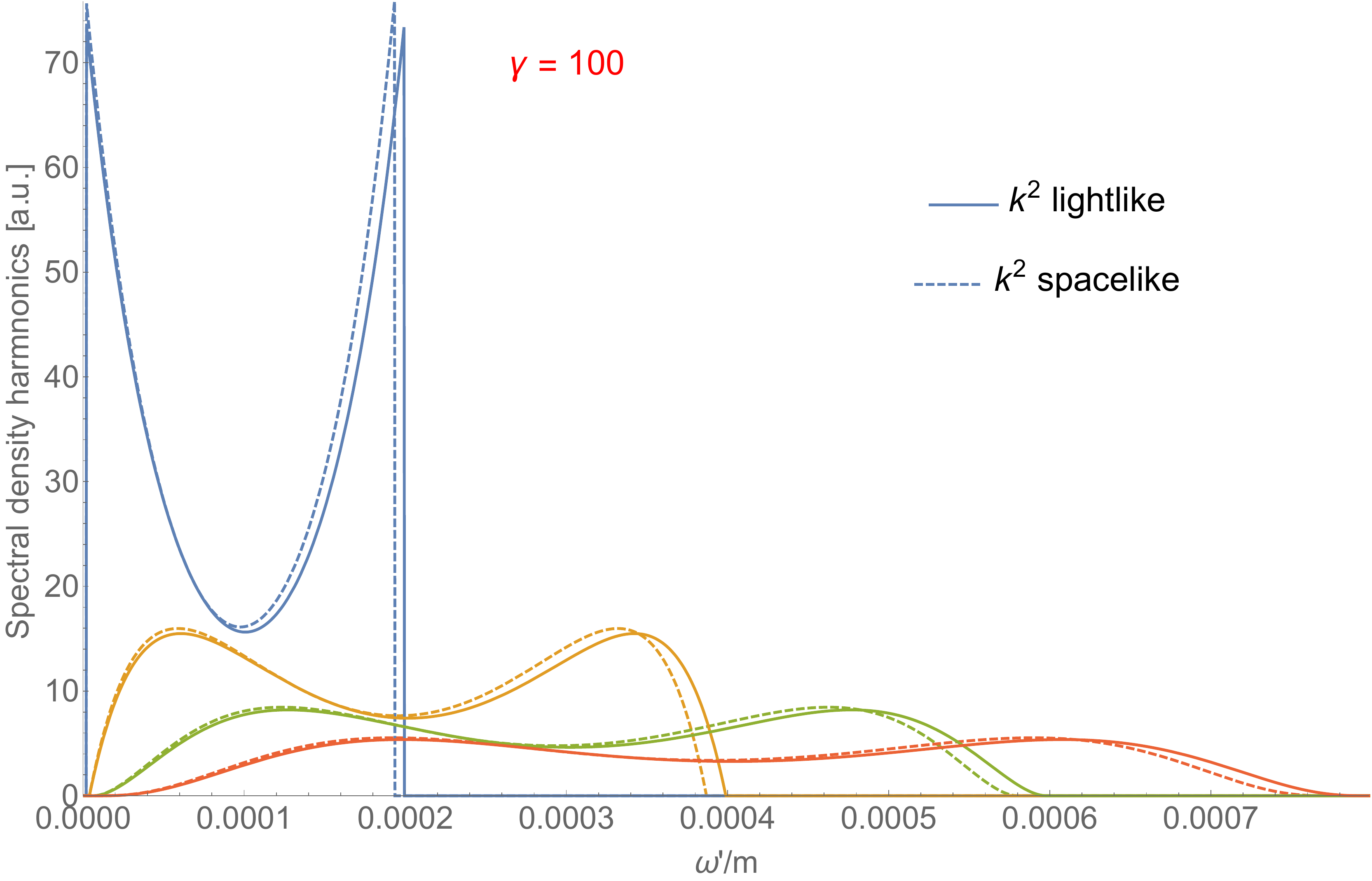}
\caption{\label{FIG:QCOMP1} Comparison of the harmonic structure in the undulator and plane wave spectra, as a function of emitted photon frequency $\omega'$, for two values of incoming electron $\gamma$. The plane wave frequency is optical, $\omega_l=2 \times 10^{-6}m$, and the undulator frequency, i.e.~the inverse of the undulator period, is fixed such that $k.p$ is the same for both fields, see (\ref{omega-omega}). As $\gamma$ increases the spectra start to overlap.}
\end{figure}

The lightlike limit $k^2\to 0$ takes $s\to1$, and by inspection of (\ref{L})-(\ref{J}) we can see 
that $s$ differentiates between the cases $k^2=0$ and $k^2<0$.  If $k^2$ is made large, so that 
$s\simeq 0$ rather than $1$, then particle dynamics and hence the emission spectrum will be very 
different in the two fields. (Further, we can take $k.p<0$ when $k^2<0$, which also suggests large 
differences because $k.p$ is always positive in the plane wave case.) However when $k^2$ is 
sufficiently small such that $s\simeq 1$, we can define an `equivalent' plane wave field for which 
the emission spectrum will almost match that of an undulator, by specifying the invariants $a_0$ and 
$k.p$. To illustrate, take $a_0=20$, $p_\mu=m\gamma(1,0,0,-\beta)_{\mu}$ and 
$k_\mu=\omega_u(0,0,0,1)_{\mu}$ for the static magnetic field. For the plane wave we take 
$k_\mu=\omega_l(1,0,0,1)_{\mu}$ and choose the frequencies such that $k.p$ is the same in both 
systems. This requires
\be\label{omega-omega}
	\omega_l = \frac{\beta}{1+\beta}\omega_u \;.
\ee
(We write equivalent `plane wave' rather than `laser' because, in contrast to (\ref{omega-omega}), 
laser frequencies are much larger than the frequencies associated with the geometry of undulators. 
For an optical laser $\omega_l\sim 1\,$eV, while for an undulator with period $\lambda_u$ of order 
$1\,$cm, the frequency is $\omega_u = 2\pi c /\lambda_u \sim 10^{-4}\,$eV.) A comparison of the 
emission spectra in the monochromatic field, for two values of $\gamma$, is shown in 
Fig.~\ref{FIG:QCOMP1}. For the lower value, $\gamma=40$, we have $s\simeq 0.87$ and there is a clear 
difference -- both the amplitudes and ranges of the spectral harmonics differ between the two cases. 
For the higher energy of $\gamma=100$ however, we have $s\simeq 0.98$ and the two spectra are 
almost in agreement. 
The assumptions of a head-on collision and a periodic field~\cite{Raicher:2013cja} give a well understood system and a testing ground for new approaches, but on the other hand represent an oversimplified special case. Once we allow for general $p_\mu$ the classical motion is described not by simple sinusoidal functions but by more involved elliptic functions. Similarly, nontrivial field envelopes increase the complexity of both the physics (e.g.~the emission spectrum) and the calculations. All of these aspects will come into play in the quantum theory, to which we now turn.

\section{First-order approaches to the quantum problem}\label{SECT:FIRST}
The problem to solve in the quantum theory is the identification of the wavefunctions which describe incoming and outgoing electrons and positrons, for use in scattering calculations. In scalar QED these wavefunctions~$\Phi$ are solutions of the Klein-Gordon equation 
\be\label{KG}
	(D^2+m^2)\Phi(x)=0 \;, \qquad D_\mu = \partial_\mu + i a_\mu(k.x) \;,
\ee
where $D_\mu$ is the background-covariant derivative, and where the gauge potential can be chosen as $eA_\mu = a_\mu$.  Current approaches to solving (\ref{KG}) mimic that used in the plane wave case, by making the ansatz $\Phi = e^{-ip.x}F(k.x)$ where $p^2=m^2$~\cite{Mendonca,Raicher:2013cja}. With this the Klein-Gordon equation reduces to
\be\label{F.PPRIME}
	k^2 F'' - 2ik.p F' + (2a.p-a^2) F = 0 \;.
\ee
For $k^2=0$ this is a first-order equation which immediately yields the Volkov solutions. For all $k^2\not=0$ the equation is second order, and there is no general solution because of the arbitrarily varying prefactor of $F$, a non-constant coefficient akin to the potential in the Schr\"odinger equation. As the difficulty here comes from the fact that the equation (\ref{F.PPRIME}) is second order, we begin by discussing a variety of approaches which seek to reduce~(\ref{F.PPRIME}) to a first-order equation.
\subsubsection{Perturbation theory and a slowly varying envelope approximation}
For plane waves with $k^2=0$, the solution to (\ref{F.PPRIME}) is~\cite{Volkov}
\be\label{SOLN:VOLKOV}
	F_\text{V}(\phi) = \exp\bigg[-i\int\limits_{-\infty}^\phi\!\ud\varphi\; u(\varphi)\bigg] \;.
\ee
If we are interested in `perturbing around' the plane wave solution, then perturbation in $k^2$ 
would seem to be a natural approach. However, in the frame where $k_\mu=(\omega_u,0,0,0)_{\mu}$ or 
$k_\mu=(0,0,0,\omega_u)_{\mu}$ for $k^2>0$ respective $k^2<0$, perturbation in $k^2$ clearly 
corresponds 
to a low-frequency expansion, whereas the Volkov solution (\ref{SOLN:VOLKOV}) makes no assumption 
about the frequency scale, and indeed $u(\phi)$ is, using (\ref{L}), nonperturbative in $k.p =  
\omega_u m \gamma$.

Of course $k^2$ has dimensions, so we need a second scale to compare against in 
order to develop a meaningful expansion in a dimensionless parameter. The approach 
of~\cite{Mendonca} is to make a slowly varying envelope approximation which, in our notation and 
made covariant, reads
\be
	| k^2 F''| \ll |k.pF'| \;.
\ee
From this, one can identify dimensionless $\epsilon = k^2/(2k.p)$ as a potential expansion 
parameter 
(the factor of 2 is for later convenience). Using perturbation in $\epsilon$ then corresponds, at 
lowest order, to dropping the double-derivative term in (\ref{F.PPRIME}). The resulting 
first-order 
equation is immediately integrable and the solution is formally identical to the Volkov solution 
(\ref{SOLN:VOLKOV}) but with $k_\mu$ spacelike or timelike, rather than lightlike:
\be\label{SOLN:mendonca}
	F_\text{pert}(x) = \exp\bigg[-i\int\limits_{-\infty}^\phi\!\ud\varphi\; u(\varphi)\bigg] \;.
\ee

\subsubsection{First-order approximation}
A different approach is given in~\cite{Raicher:2015ara,Raicher:2016bbx}, but only for a specific field shape (monochromatic, and circular polarisation, as above). The idea is again to replace (\ref{KG}) with a soluble first-order equation. The choice of this equation is motivated by taking its derivative, and showing that it reproduces (\ref{KG}) up to terms small in some parameter. In~\cite{Raicher:2015ara,Raicher:2016bbx} the parameter is, 
\be\label{Raicher:delta}
	\delta \sim \frac{a_0 m k^2 \sqrt{-l_j.p \, l_j.p}}{k.p^2+a_0^2 m^2 k^2} = \frac{a_0 m 
|p_\LCperp|}{p_0^2+a_0^2 m^2} \;,
\ee
where the last identity holds in the frame where $k_\mu=(\omega_u,0,0,0)_{\mu}$, introducing
$p_\LCperp=\{p_1,p_2\}$. A small $\delta$ means that the particle's initial 
transverse momentum should be much smaller than the total energy, and the field strength. This approach has the potential to be applied to other field shapes.

\subsubsection{Reduction of order}
We present now an alternative method which combines the perturbative approach above with that in~\cite{Raicher:2015ara,Raicher:2016bbx}: we again look for an `effective' first-order equation, but related to an expansion in the small parameter $\epsilon$. Observe that dividing (\ref{F.PPRIME}) by $2k.p$ makes the coefficients dimensionless, 
and the equation becomes
\be \label{F.PPRIME2}
     i F' = u(\phi) F  + \epsilon F'' \;.
\ee
If $\epsilon$ is a small parameter then (\ref{F.PPRIME2}) constitutes a typical example of singular perturbation theory~\cite{Bender:1978}, in which the highest derivative term is multiplied by the small parameter. This situation is familiar from the study of radiation reaction in strong fields, where the third-derivative term in the Lorentz-Abraham-Dirac (LAD) equation~\cite{L,A,D} is a singular perturbation in the same sense as encountered here. We therefore apply the Landau-Lifshitz approach~\cite{Landau:1982dva} to our problem. This means using recursion: taking another derivative of, and employing, (\ref{F.PPRIME2}) yields
\be \label{ITER2}
	 F'' = (-i u' -u^2)F - \epsilon u F'' - i\epsilon F''' \;.
\ee
Plugging this into (\ref{F.PPRIME2}) and discarding terms of order $\epsilon^2$ and higher again 
leaves a first-order equation,
\be \label{ITER3}
  F' = (-i u + i \epsilon u^2 - \epsilon u') \, F \; .
\ee
(This approach can be extended directly to include higher orders of $\epsilon$ in (\ref{ITER3}).) 
This is a non-singular perturbation of the plane-wave equation which is still reproduced in the 
limit $\epsilon \to 0$. The solution to~(\ref{ITER3}) is
\be \label{F.PERT}
  F_\text{RO} = \exp \bigg[-i \int\limits_{-\infty}^\phi \!\ud\phi' \, ( u - \epsilon u^2)- 
\epsilon u(\phi) \bigg] \;.
\ee
For $\epsilon=0$ we recover (\ref{SOLN:mendonca}) for any value of $k^2$ (so, again, this is not an 
expansion `around' the plane wave case).

Using `reduction of order' (RO) we thus obtain a wavefunction which is no more or less complicated 
than the Volkov solution and which, like Volkov, applies for any field shape. We refer to this as a 
`partial resummation' of the perturbative series because although we have thrown away terms of order 
$\epsilon^2$ from the equation (\ref{ITER3}), the solution (\ref{F.PERT}) contains all orders in 
$\epsilon$.

RO~is not completely general because, for $k$ spacelike, $k.p$ and therefore $\epsilon$ can be of any sign and size (in the `undulator frame', for example, $\epsilon=\omega_u/p_z$). When $\epsilon$ is not small, the arguments leading to RO~do not hold. 

\section{Second-order approaches to the quantum problem}\label{SECT:SECOND}
The approaches above try to solve the problem at hand based on experience of the plane wave case, 
hence the focus on eliminating the second derivative in (\ref{F.PPRIME}). It is however not obvious 
that this is the best approach to take. For example, the dimension of the solution space changes 
from two to one when going from second to first-order, which can correspond to the decoupling of 
states from a theory~\cite{Heinzl:2007ca}; it is not clear \emph{apriori} that these states should 
be discarded.

Consider then the Schr\"odinger equation. This is a well-understood second order equation in 
quantum 
physics where the second derivative term is retained despite being multiplied by a parameter 
(Planck's constant) considered `small' in the semi-classical limit. In this section we 
therefore rewrite the Klein Gordon equation (\ref{KG}) as a Schr\"odinger equation, and use 
intuition from quantum mechanics to gain insight into its solutions and into finding 
physically-motivated approximations. 

Ultimately we are still interested in solutions with scattering boundary conditions, i.e.~which go 
like $e^{-ip.x}$ asymptotically, so that the solution is free far from the field, but we make an 
ansatz
\be\label{decomp}
	\Phi(x) = e^{-i\tilde{p}.x}G(k.x) \;, \quad \tilde{p}_\mu := p_\mu - \frac{k.p}{k^2}k_\mu 
\;.
\ee
Compared to the ansatz in Sect.~\ref{SECT:FIRST}, this simply moves all $\phi$-dependence in 
$\Phi(x)$ into the unknown function $G$.  (This new ansatz corresponds to transforming 
(\ref{F.PPRIME2}) to ``normal form''~\cite{Polyanin:1994} using a Galilei boost to remove the first 
derivative term.) The Klein-Gordon equation then becomes
\be\label{KGett}
	k^2 \frac{\ud^2G(\phi)}{\ud\phi^2} + (2a.p-a^2) G(\phi) = \big(\tilde{p}^2-m^2\big)G(\phi) 
\;,
\ee
which we recognise as a Schr\"odinger equation for $G$,
\be\label{SCHRO}
	-\frac{\hbar^2}{2} \frac{\ud^2G(\phi)}{\ud\phi^2} + V(\phi) G(\phi) = \mathcal{E} G(\phi) 
\;,
\ee
and the first task is to map (\ref{KGett}) to (\ref{SCHRO}) by identifying the form and relative  
sizes of the potential $V$, the energy eigenvalue $\mathcal{E}$ and the analogue of $\hbar$. We 
will 
now see through three examples that this identification and the corresponding approximate solutions 
to (\ref{SCHRO}) depend sensitively on both the parameters and the background field shape. 

\subsection{Example 1: over the barrier}
For a general scattering amplitude we need both incoming and outgoing states. Consider first the 
case of incoming particles, incident on the field (\ref{FIELD}). We have $k^2<0$ and $k_\mu$ 
contains the typical frequency scale of the background. Write the field as $a_\mu = m a_0 
\hat{f}_\mu$ where $a_0$ is the amplitude and $\hat f$ is of order unity. In addition to taking 
$k^2$ small, the typical case of interest for incoming particles is a nearly head-on collision, so 
that $k.p/\sqrt{-k^2}\gg p_\LCperp$. We also assume $p_\LCperp\ll a_0 m$ i.e.~that the initial 
transverse momentum is much smaller than the typical transverse momentum acquired in the field. Then 
taking (\ref{KGett}) and dividing by $m^2a_0^2$ we identify
\be\label{VSIM}
	V\sim \hat{f}^2 \sim 1 \;,
\ee
of order unity and
\be
	-\frac{\hbar^2}{2} = \frac{k^2}{a_0^2m^2} \ll 1\;,
\ee
so that small $\hbar$ corresponds to the semiclassical limit. We begin by assuming that the 
dimensionless field strength $a_0$ is much lower than the particle energy $\gamma$, so that
\be\label{ESIM}
	\mathcal{E} \sim \frac{\gamma^2}{a_0^2} \gg 1 \;.
\ee
Comparing (\ref{ESIM}) and (\ref{VSIM}) shows that this situation is analogous to over the barrier 
scattering in quantum mechanics. As such the natural approximation with which to solve (\ref{SCHRO}) 
is semiclassical WKB~\cite{Bender:1978}:
\be\label{WKB}
	G(\phi) \sim \bigg(\frac{1}{\mathcal{E}-V(\phi)}\bigg)^{\frac{1}{4}}\exp\bigg[\pm 
\frac{i}{\hbar} \int\limits^\phi\sqrt{2(\mathcal{E}-V)}\bigg] \;.
\ee
The sign may be fixed either by boundary conditions on $k.p$ at $\phi\to\pm\infty$, or by the 
$a_0\to 0$ limit. The WKB wavefunction is based on making a free-field ansatz, and works well in the 
case that the particle is almost free due to its high energy~\cite{Bender:1978}, independent of the 
precise form of the potential $V$.
\subsection{Example 2: wide-angle scattering/under the barrier}\label{spreading}
Quantum effects can cause wide-angle particle scattering, where classical motion 
cannot~\cite{Green:2013sla}. In this case the assumptions which are natural for incoming particle 
states (high energy, nearly head-on) break down, and one must instead consider large transverse 
momenta.  In this case the Schr\"odinger equation may describe, in contrast to the above, a 
below-the-barrier scattering problem.

To illustrate we use the Sauter pulse with gauge potential ($k.l=0$ as 
in~(\ref{FIELD})) 
\be
	a_\mu(\phi) = a_0 m l_\mu \big(\tanh(\phi)+1\big) \equiv a_0 m l_\mu \hat{f}(\phi)\;.
\ee
For $k^2=0$ this would correspond to a short, subcycle laser pulse. We again write 
$\Phi=e^{-i\tilde{p}.x}G(\phi)$, with $\tilde{p}$ as above. The field can give a classical particle 
a transverse momentum of order $a_0 m$, so we measure $p_\LCperp$ in these units, writing $p_\LCperp 
= a_0m \kappa l_\LCperp$.  With this the Klein-Gordon equation again reduces to a Schr\"odinger 
equation
\be\label{G-UT-1}
	\frac{k^2}{a_0^2m^2}G'' + (\hat{f}^2 -2 \kappa \hat{f})G = \frac{\tilde{p}^2-m^2}{a_0^2m^2}G 
\;,
\ee
\begin{figure}[t!]
	\includegraphics[width=0.9\columnwidth]{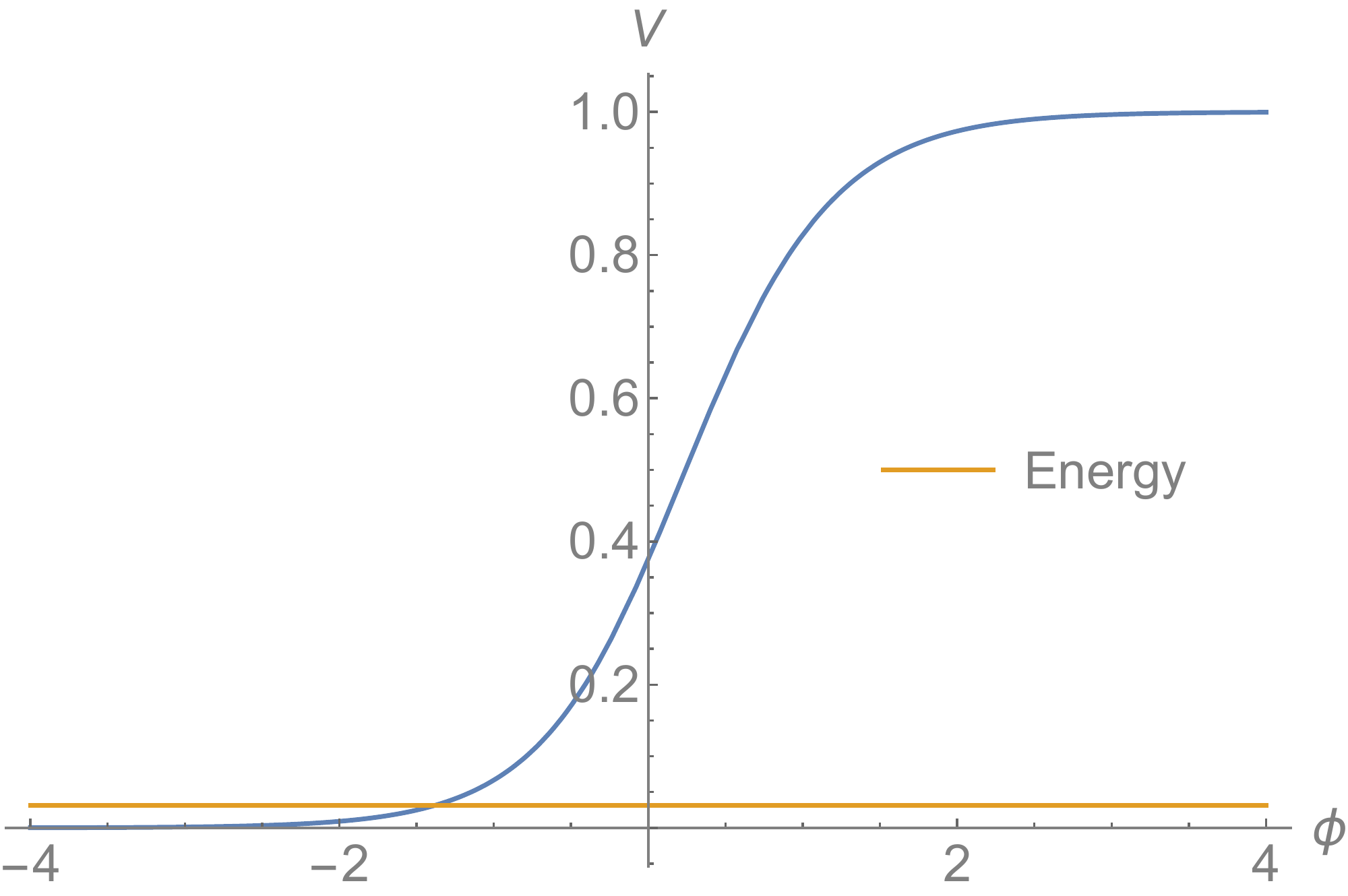}
	\includegraphics[width=\columnwidth]{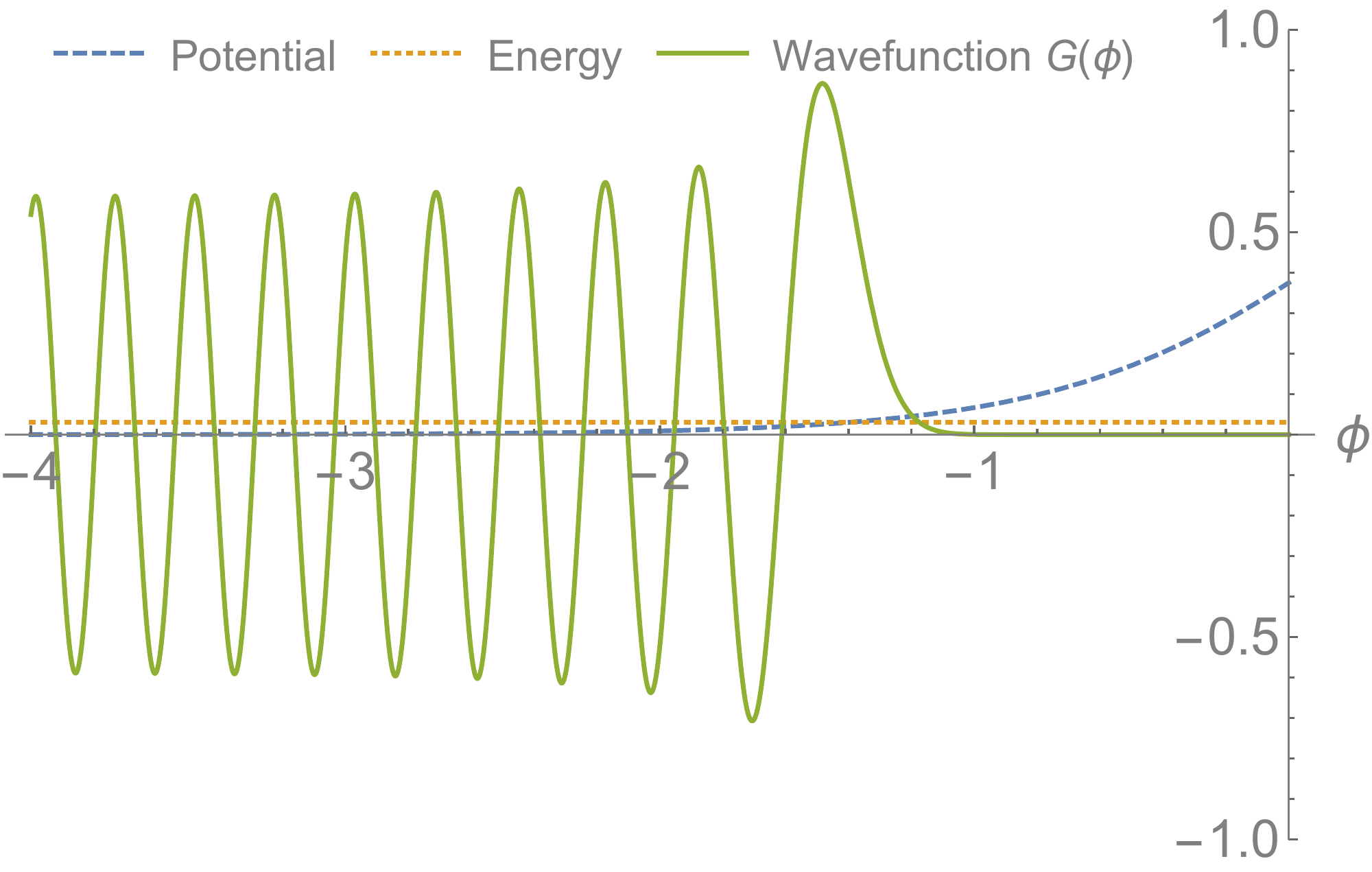}
	\caption{\label{FIG:STEG1} \textit{Upper panel:} normalised potential function $V$ for 
$\kappa=-1$ in the Sauter field, and energy eigenvalue $\mathcal{E}=1/32$. \textit{Lower panel:} 
U.WKB wavefunction for $\hbar=1/100$, shown together with the potential and energy eigenvalue. The 
wavefunction is exponentially damped under the barrier, i.e.~when $\mathcal{E}<V$, and oscillatory 
when $\mathcal{E}>V$, above the potential. The numerical solution of (\ref{G-UT-1}) is 
indistinguishable from the U.WKB approximation on the scale shown.}
\end{figure}

\noindent and we read off $\hbar$, $V$ and $\mathcal{E}$ as above (possibly after dividing by some 
$\kappa$-dependent constant in order to normalise the potential to $|V|\leq1$). To impose the 
condition of wide-angle scattering we take $\kappa$ of order unity, and $\tilde{p}^2\simeq m^2$, 
which corresponds to the longitudinal momentum $\propto k.p$ being much smaller than the transverse 
momentum. Thus the eigenvalue now obeys $\mathcal{E}\gtrsim 0$, as opposed to $\mathcal{E}\gg1$ in 
the case of near-forward collisions considered above. An example is shown in Fig.~\ref{FIG:STEG1} 
for $\kappa=-1$. As $\phi$ varies, $V(\phi)-\mathcal{E}$ can change sign, which means we have a 
barrier penetration, or tunnelling problem. This suggests using a \textit{uniform} WKB (``U.WKB'') 
ansatz for the wavefunction, of which ordinary WKB is a special case. We make the U.WKB 
ansatz~\cite{Cherry,MillerGood,Miller}
\be\label{U-WKB}
	G(\phi) = \frac{1}{\sqrt{\varphi'(\phi)}} 
\text{Ai}\bigg(\frac{2^{1/3}}{\hbar^{2/3}}\varphi(\phi)\bigg) \;,
\ee
and expand $\varphi$ as a series in $\hbar^2$. (This ansatz, used by Sauter to study the behaviour of an electron in a homogeneous field~\cite{Sauter31}, gives the exact solution in a linear potential.) To lowest order the equation to solve is
\be
	\varphi{\varphi'}^2 = V-\mathcal{E} \;,
\ee
which has the two solutions
\be\label{VARPHI-SOLN}
	\varphi_\LCpm(\phi) = \bigg(\pm\frac{3}{2}\int\limits^\phi_0 \ud z 
\sqrt{(V-\mathcal{E})}\bigg)^{2/3} \;.
\ee
In our case the integral in (\ref{VARPHI-SOLN}) can be performed analytically, but is an unrevealing 
combination of hyperbolic functions. This gives two independent solutions to (\ref{G-UT-1}). 
(Equivalently, one can use only $\phi_\LCp$ but include both Ai and Bi terms in the ansatz for 
$G$~\cite{DLMF:AIRY}.) Demanding that the wavefunction is continuous at the turning point and that 
its amplitude does not diverge asymptotically then determines the solution.

The result is plotted in Fig.~\ref{FIG:STEG1}. The wavefunction is real, and is indistinguishable 
from a numerical solution of the ODE (\ref{G-UT-1}) on the scales shown, demonstrating the accuracy 
of the approach. Importantly, the U.WKB wavefunction clearly reproduces the expected physics: when 
$\mathcal{E}>V(\phi)$ above the barrier and the wavefunction is oscillatory, but when 
$\mathcal{E}<V(\phi)$ under the barrier and the wavefunction is exponentially damped. In 
Sect.~\ref{RO2} we will compare these results with those obtained from the first-order 
approximations.

\subsection{Example 3: periodic fields and the Mathieu equation.}\label{SECT:MATHIEU}
For our final example we return again to the case of monochromatic, circularly polarised fields. The 
Klein-Gordon equation for $G$ then becomes equivalent to the Mathieu 
equation~\cite{DLMF,Muller-Kirsten:2012wla}
\be\label{MATHIEU}
	\frac{\ud^2G}{\ud y^2} - 2 Q \cos (2 y)G = -A G  \;,
\ee
with the identifications $\phi=2y$ and~\cite{Cronstrom:1977,Becker:1977}
\be\label{a-and-q}
	A := \frac{4}{k^2} \bigg(\frac{k.p^2}{k^2}+ a_0^2 m^2\bigg) \;, \quad Q = -\frac{4a_0 m 
|p_\LCperp|}{k^2} \;.
\ee
Despite the apparent simplicity of the classical theory, the quantum theory exhibits an intricate 
nonperturbative structure for $k^2\not=0$~\cite{Bender:1978,Dunne:2016qix}. The $A$--$Q$ parameter 
space of solutions can be divided into `bands' and `gaps'; for parameters in the gaps, the solutions 
to (\ref{MATHIEU}) increase exponentially with $y$ (or $\phi$) and cannot be normalised, so must be 
discarded as unphysical~\cite{Cronstrom:1977,Becker:1977}. For a recent discussion of the band 
structure in the language of resurgence, see~\cite{Dunne:2016qix} and references therein.
  
To map the Mathieu equation to the Schr\"odinger equation (\ref{SCHRO}) note that $u(\phi)$ contains 
now a constant term, which we move into the eigenvalue, identifying
\be\begin{split}
	\frac{\hbar^2}{2} = \frac{2}{Q} \;, \quad V=\cos\phi\;, \quad \mathcal{E} &= \frac{A}{2Q} 
\;.
\end{split}
\ee
These identifications differ from those used in the examples above, illustrating the dependence of 
the system, approach, and solutions on the form of the potential. To understand how the parameters 
affect the physics, it is convenient to focus on a particular observable, for which we choose the 
quasi-momentum. Classically, this is just the cycled-averaged particle momentum. Quantum 
mechanically, it can be identified as the cycle-average of the exponent of the wavefunction. The 
frequencies of photons emitted by an electron in our field are determined by the conservation of 
quasi-momentum according to 
\be\label{bevarat}
	q_\mu + n k_\mu = q'_\mu + k'_\mu \;,
\ee
where $q$ ($q'$) is the quasi-momentum of the incoming (outgoing) electron and $k'$ is the emitted 
photon momentum; (\ref{bevarat}) is familiar from the plane wave case~\cite{LL-QED}. In the 
classical limit, and for e.g.~a head-on collision, (\ref{bevarat}) becomes equivalent to the 
support 
of the delta function in (\ref{D3N}).

\begin{figure}[t!]
\includegraphics[width=\columnwidth]{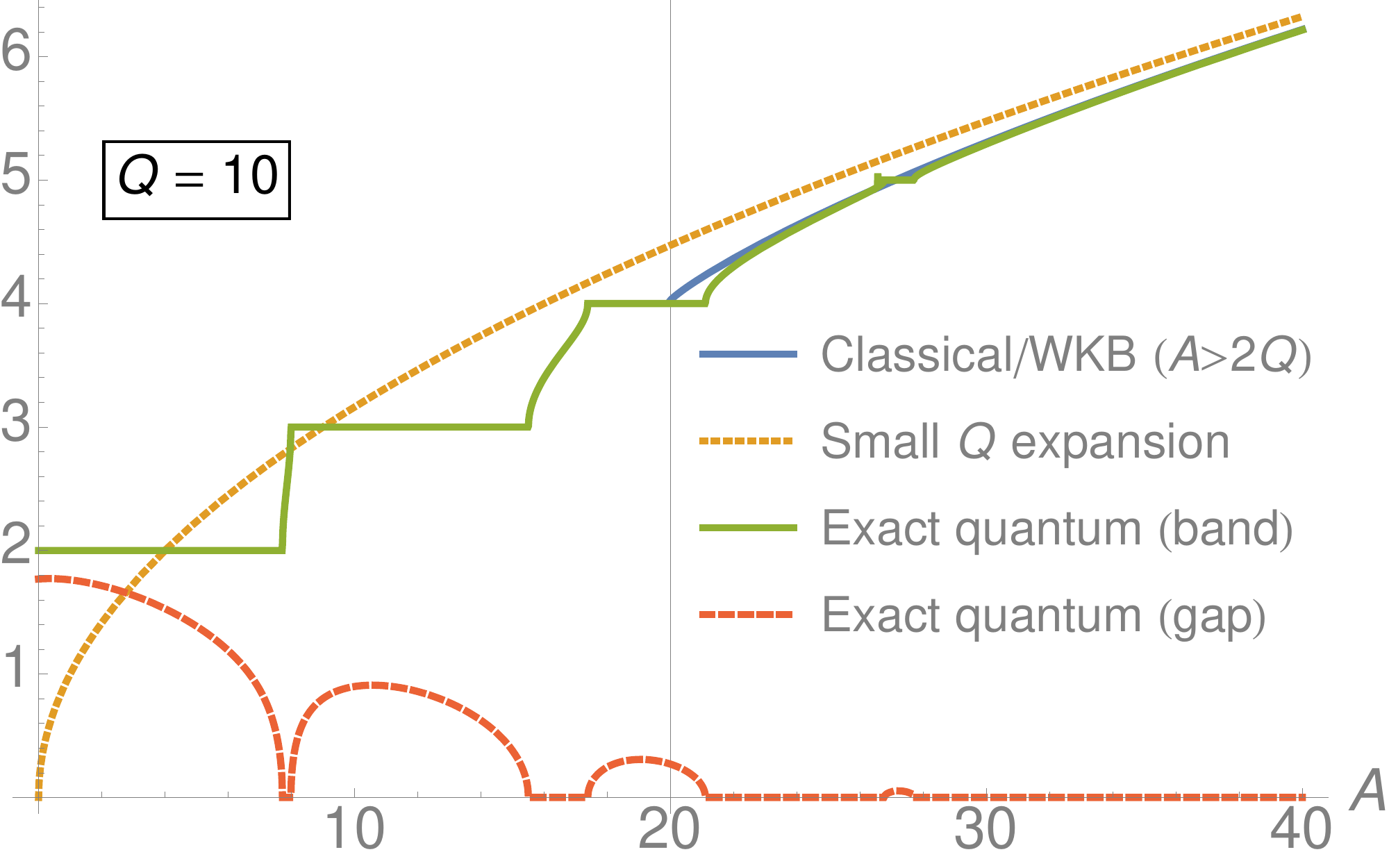}
\caption{\label{FIG:QUASI-Q-10} Comparison of the longitudinal quasi-momentum component in 
(\ref{Q-KL}). The real and imaginary parts of the exact quantum solution are shown with green/solid 
and orange/dashed lines respectively. Blue/solid: WKB (equal to exact classical). Yellow/dotted: 
small $Q$ expansion. Here $Q$~is large, i.e.~$\hbar$ is small. The vertical grey line shows the 
position of the barrier top, $A=2Q$.  For $A>2Q$, above the barrier, the exact quantum expression 
is 
well estimated by the semiclassical WKB result, and the small $Q$ expansion is not as accurate.  
For 
$A<2Q$, under the barrier, the band and gap structure is clearly visible. Neither approximation 
sees the band/gap structure, either over or under the barrier, where the gaps are respectively 
broad 
and narrow~\cite{Dunne:2016qix}.}
\end{figure}
\begin{figure}[t!]
\includegraphics[width=\columnwidth]{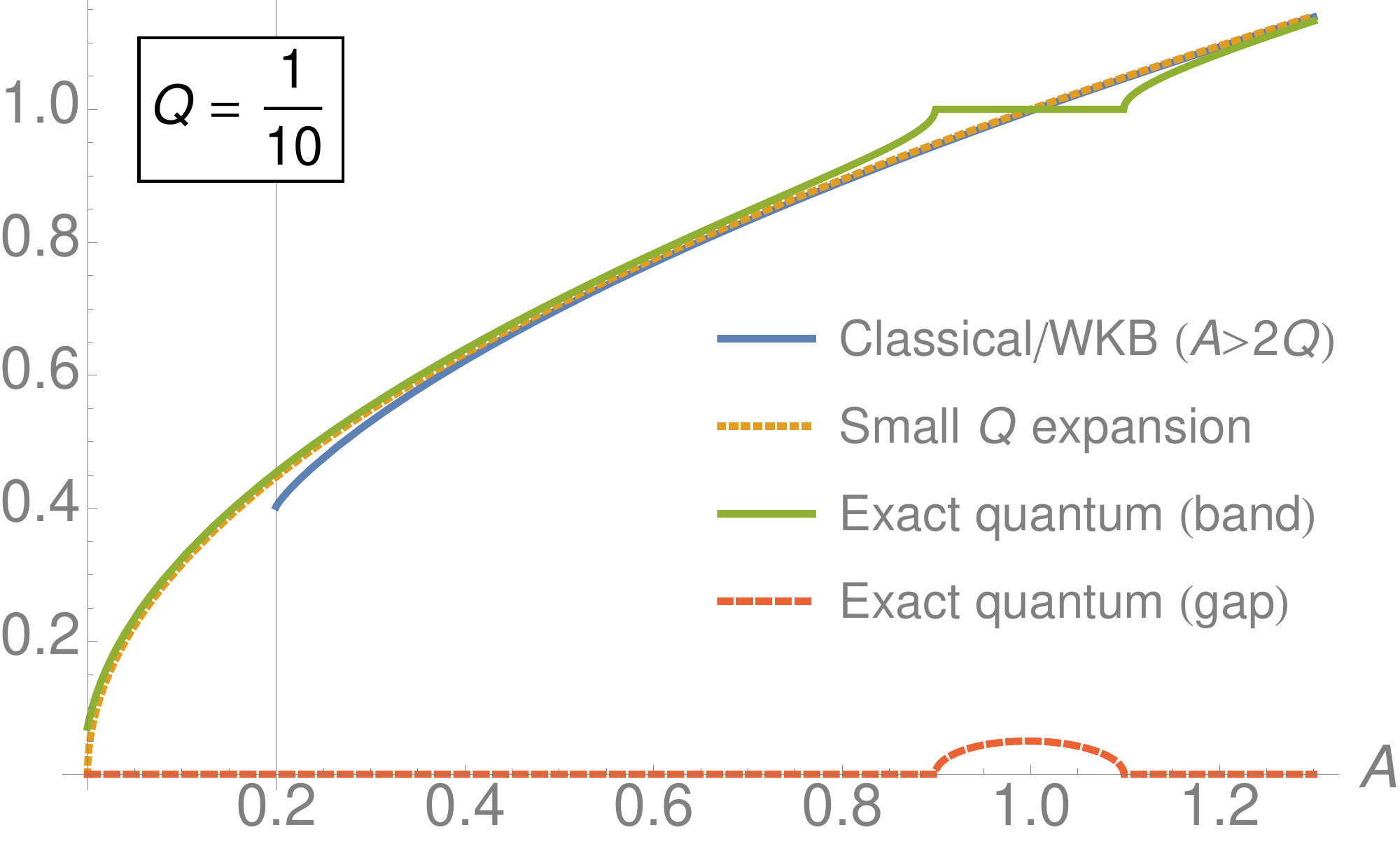}
\caption{\label{FIG:QUASI-Q-TENTH} The quasi-momentum as in Fig.~\ref{FIG:QUASI-Q-10}, but for $Q$ 
small, i.e.~$\hbar$ large. The small $Q$ (strong coupling) expansion is accurate both above and 
below the barrier. Even above the barrier, $A\gtrsim 2Q$, the small $Q$ expansion provides a better 
estimate than the WKB result, as $\hbar$ is not small.}
\end{figure}

The exact form of the quasi-momentum in the quantum case is 
known~\cite{Cronstrom:1977,Becker:1977}: 
using the notation of~\cite{Becker:1977} we have
\be\label{Q-K}
	q_\mu = \tilde{p}_\mu -\frac{1}{2}\text{sign}(k.p)\nu(A,Q) k_\mu \;,
\ee
where $\nu(A,Q)$ is the `Mathieu characteristic exponent'~\cite{DLMF}. The bands, i.e.~the spaces 
of 
physical solutions, are defined by the condition $\text{Im}(\nu) 
=0$~\cite{Cronstrom:1977,Becker:1977,DLMF}. The exact result (\ref{Q-K}) can be compared with the 
approximations above. The WKB wavefunction gives the quasi-momentum as (\ref{Q-K}) but with
\be\label{Q-KL}
	\nu(A,Q) \to  \sqrt{A - 2 Q} \; E\bigg(\frac{-4Q}{A-2Q}\bigg)\;,
\ee
in which $E$ is the complete Elliptic integral of the second kind. The exact result and WKB 
approximation in (\ref{Q-KL}) are plotted and compared in Fig.'s~\ref{FIG:QUASI-Q-10} 
and~\ref{FIG:QUASI-Q-TENTH}.  If $Q$ is large, so that $\hbar$ is small, we are in a semiclassical 
regime where WKB applies. The WKB wavefunction gives good agreement in this case, for $A>2Q$, see 
Fig.~\ref{FIG:QUASI-Q-10}. Observe that $A>2Q$ precisely when $\mathcal{E}>1$ so that the energy 
eigenvalue 
lies above the potential.  For $A<2Q$ we are within the potential, and there is a rich structure of 
bands and gaps even when $A\ll Q$ because of quantum effects from $\hbar$. However, if $Q$ is 
small, 
then $\hbar$ is large, which is a strong coupling limit~\cite{Dunne:2016qix}. In this case a direct 
small-$Q$ approximation~\cite{Becker:1977,DLMF},
	\be
		\nu(A,Q)\simeq \sqrt{A} \;,
	\ee
gives a better approximation both above and below the barrier than the WKB approximation, as shown 
in Fig.~\ref{FIG:QUASI-Q-TENTH}.

\section{Comparison of first and second order approaches}\label{SECT:COMP}
%
Having built up some intuition, we re-analyse the first-order approaches discussed in 
Sect.~\ref{SECT:FIRST}.

\subsection{Reduction of order revisited}\label{RO2}
We begin by rewriting the RO equation (\ref{F.PPRIME}) in the second order, Schr\"odinger equation 
notation of Sect.~\ref{SECT:SECOND}: 
\be\label{F.NYTT.1}
	-\frac{\hbar^2}{2}F'' \mp i \sqrt{2\hbar^2\mathcal{E}}F' + V F = 0 \;,
\ee
where the sign is minus that of $k.p$. It is easily confirmed that a straightforward perturbative 
expansion in $\hbar$ gives only the trivial solution $F=0$; this is consistent with the results in 
Sect.~\ref{SECT:SECOND} where we saw that the leading behaviour of the wavefunctions is 
non-perturbative in $\hbar$. Each derivative in (\ref{F.NYTT.1}) comes with a factor of $\hbar$; 
temporarily scaling this out by sending $\phi \to \phi/\sqrt{2 \mathcal{E}\hbar^2}$ will allow us to 
better compare the relative sizes of the first and second derivative terms. Writing a dot for a 
derivative with respect to the new variable, (\ref{F.NYTT.1}) becomes
\be\label{F.OMSKRIVNING}
	-\frac{1}{4\mathcal{E}}\ddot{F} \pm i \dot{F} + V F = 0 \;.
\ee
Recalling that $V$ is of order unity, the only candidate small parameter is $1/\mathcal{E}$. Hence 
recursion in the second-derivative term corresponds to a large $\mathcal{E}$ expansion. This is 
confirmed by also rewriting the RO~wavefunction (\ref{F.PERT}) in the Schr\"odinger equation 
notation (returning to the usual $\phi$ variable),
\be \label{F.PERT.G}
  F_\text{RO}(\phi) = \exp \bigg[ \pm i 
\sqrt{\frac{1}{2\mathcal{E}\hbar^2}}\int\limits_{-\infty}^\phi \!\ud\phi' \, (V + 
\frac{1}{4\mathcal{E}}V^2 ) +\frac{V(\phi)}{4\mathcal{E}}\bigg] \;,
\ee
and observing that $\Phi_\text{RO}$ is the third order expansion of the exponent of the WKB solution 
(\ref{WKB}) in powers of $1/\sqrt{\mathcal{E}}$, in the limit that $\mathcal{E}$ is very much 
greater than $V$. Expanding the square root in (\ref{WKB}) gives the phase needed to convert from 
$G$ to $F$, and the terms under the integral in (\ref{F.PERT.G}) and re-exponentiating the prefactor 
in (\ref{WKB}) gives a term $-\tfrac{1}{4}\log (1- V/\mathcal{E})$ in the exponent, the lowest order 
expansion of which gives the final term in~(\ref{F.PERT.G}). The expansion holds when the energy is 
far above the potential, $\mathcal{E}\gg1$, therefore RO is a high-energy approximation. This is 
consistent with the perturbative result obtained by dropping the double derivative term: the lowest 
order perturbative wavefunction~(\ref{SOLN:mendonca}) is a phase, as expected for above the barrier 
wavefunctions, but which can never describe barrier penetration (or potential well/bound 
state) problems where the solutions are asymptotically damped.

\begin{figure}[t]
	\includegraphics[width=\columnwidth]{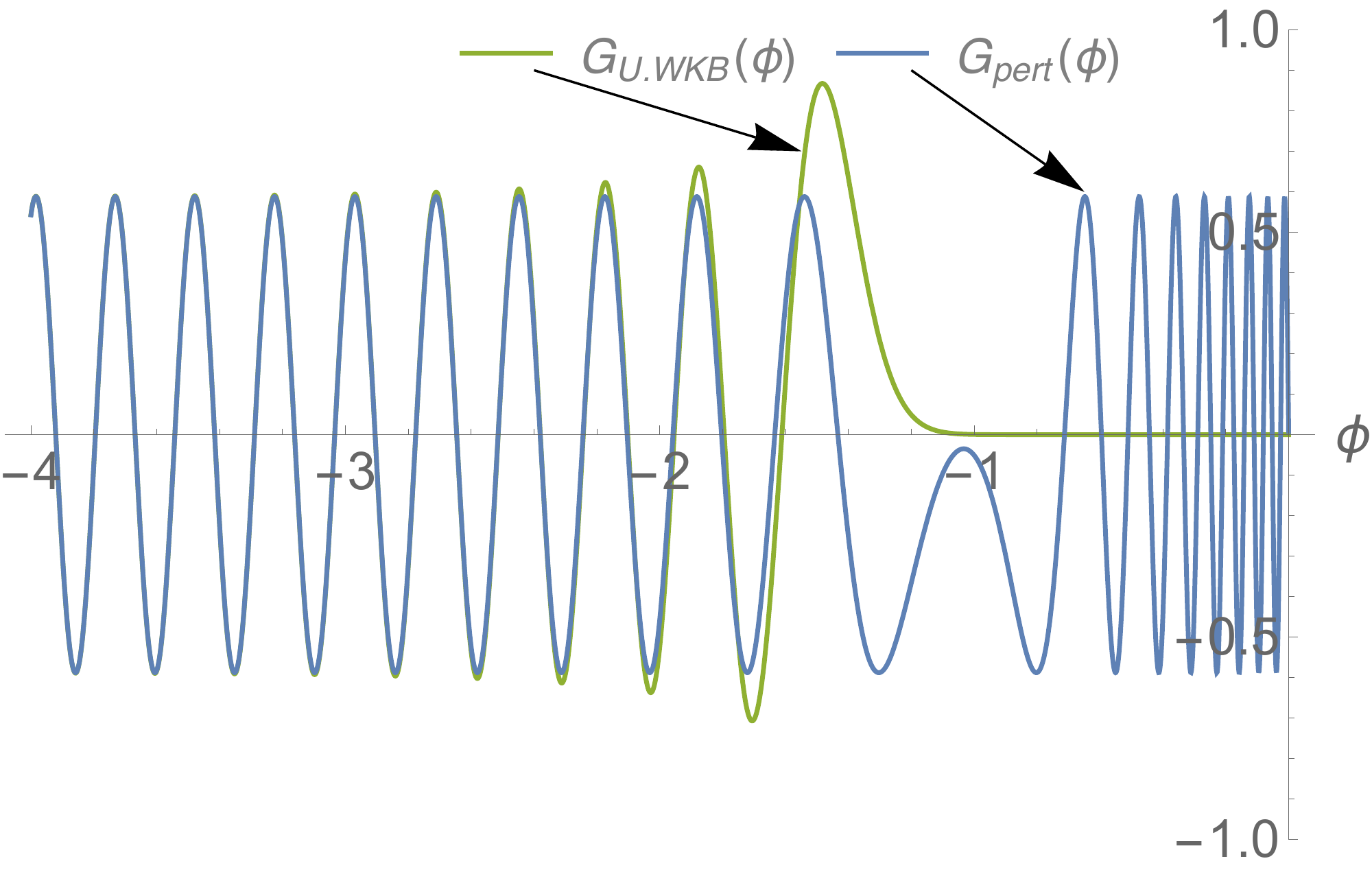}
	\includegraphics[width=\columnwidth]{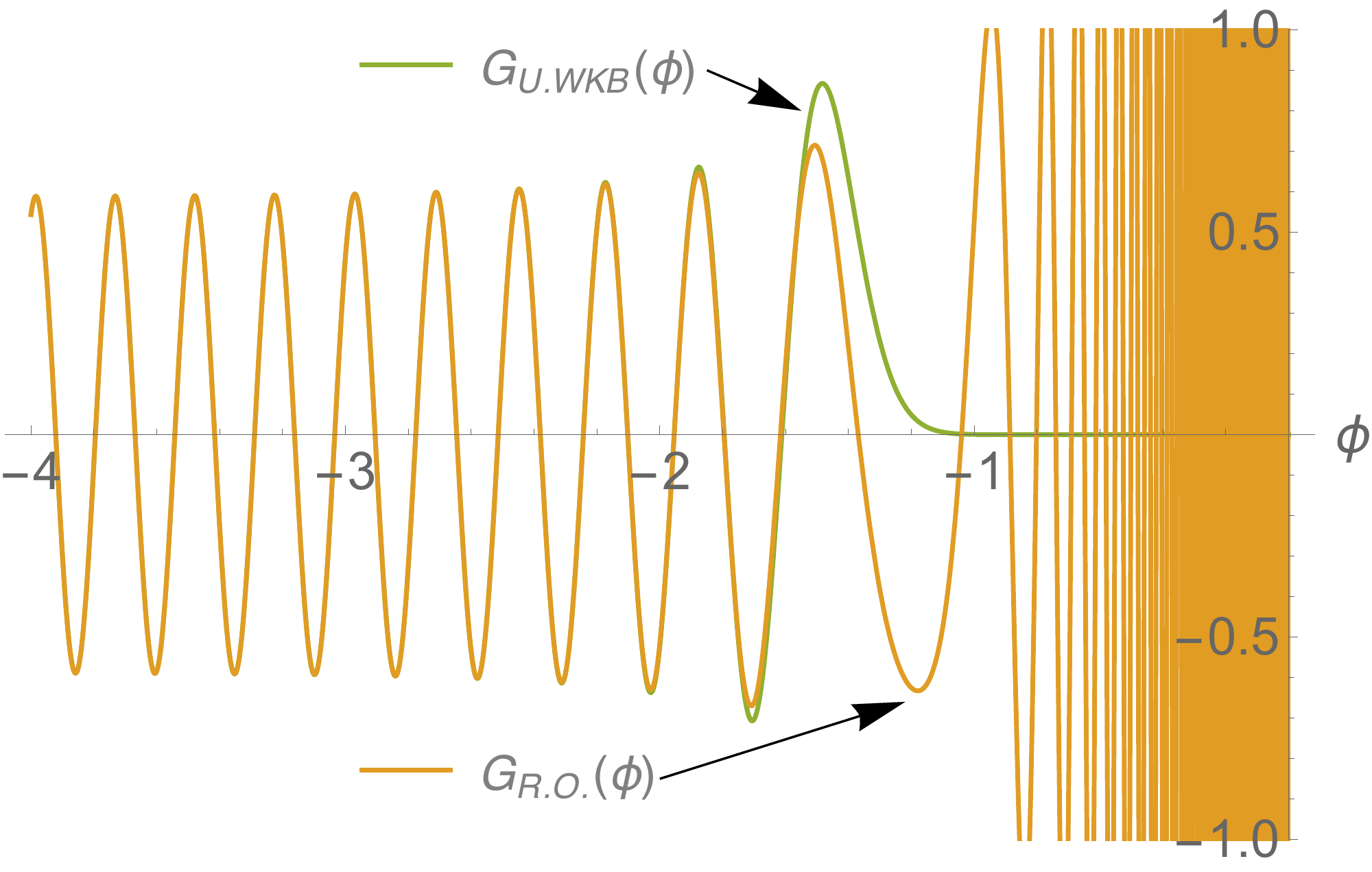}
	\caption{\label{FIG:COMP1} Comparison of the U.WKB, perturbative (top panel) and RO (bottom 
panel) wavefunctions, for the example of Fig.~\ref{FIG:STEG1}, which corresponds to the expansion 
parameter of the perturbative and RO~approaches being $\epsilon=\pm 1/50$. The RO~wavefunction more 
closely tracks the U.WKB wavefunction as we approach the turning point, but the difference is small. 
Under the barrier, neither the 
perturbative nor RO~wavefunctions fall off as expected. This is consistent with those 
approximations begin essentially high-energy.}
\end{figure}

\begin{figure*}[t]
	\includegraphics[width=0.3\textwidth]{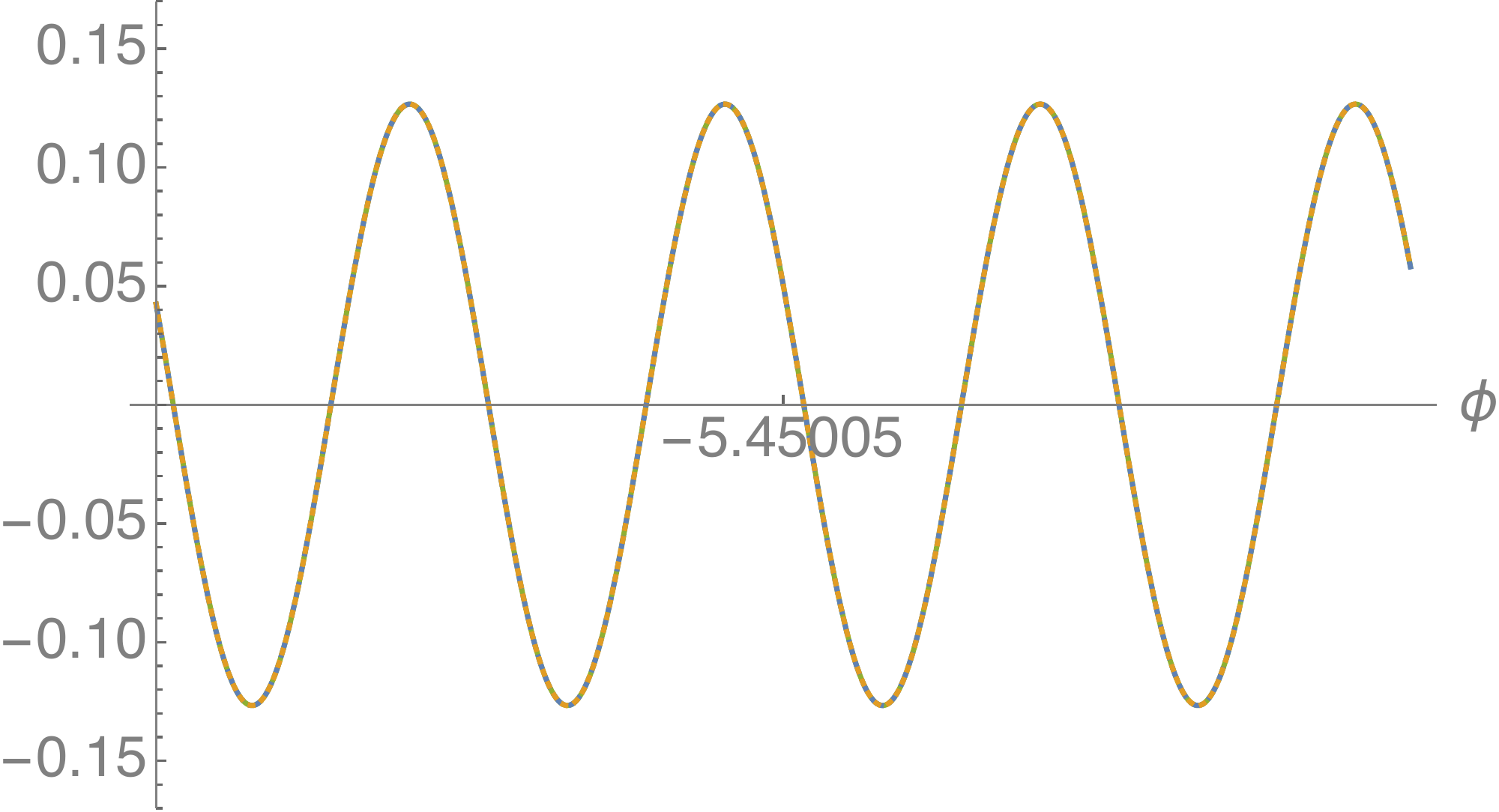}
	\includegraphics[width=0.3\textwidth]{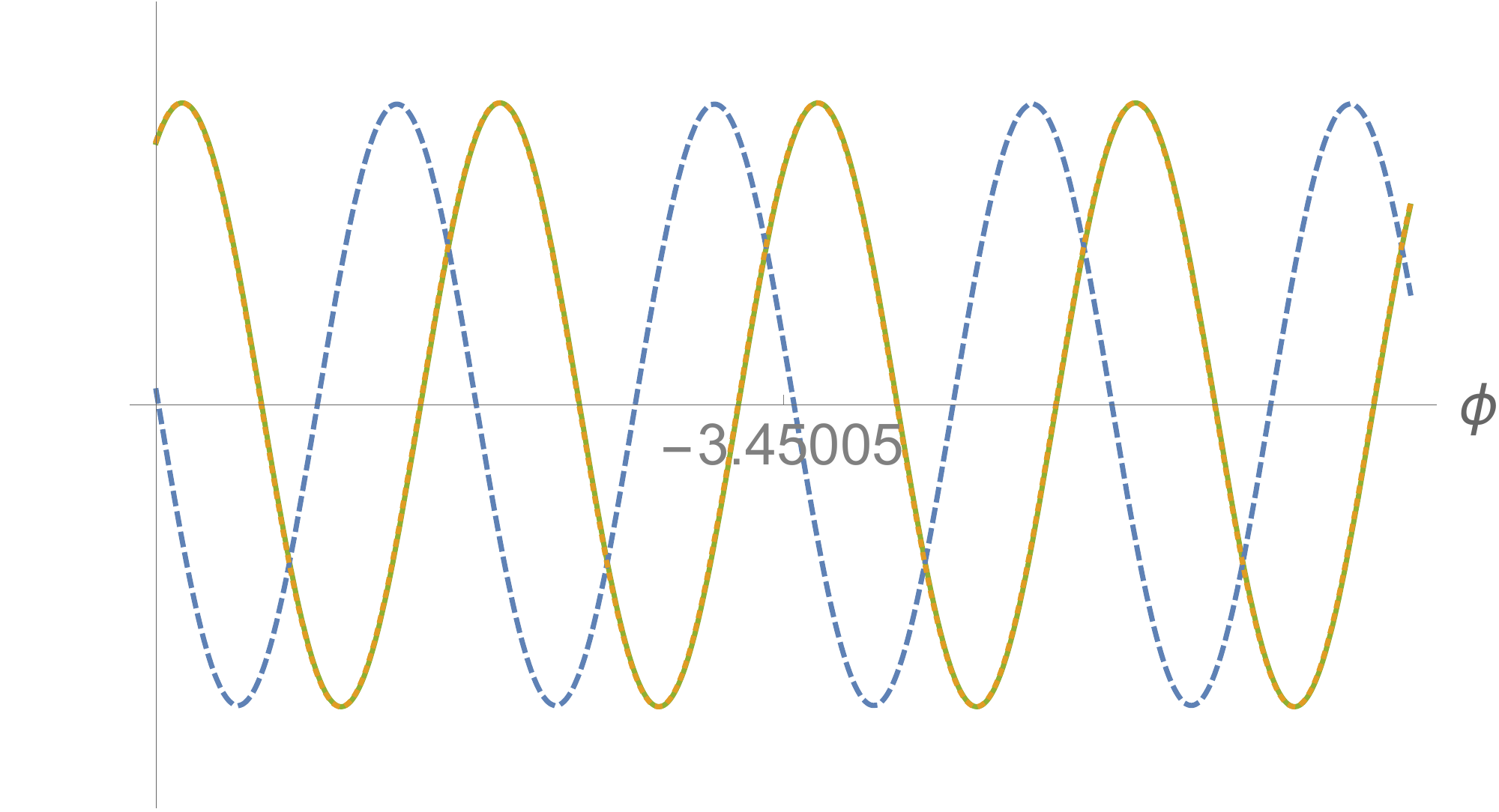} \\
	\includegraphics[width=0.3\textwidth]{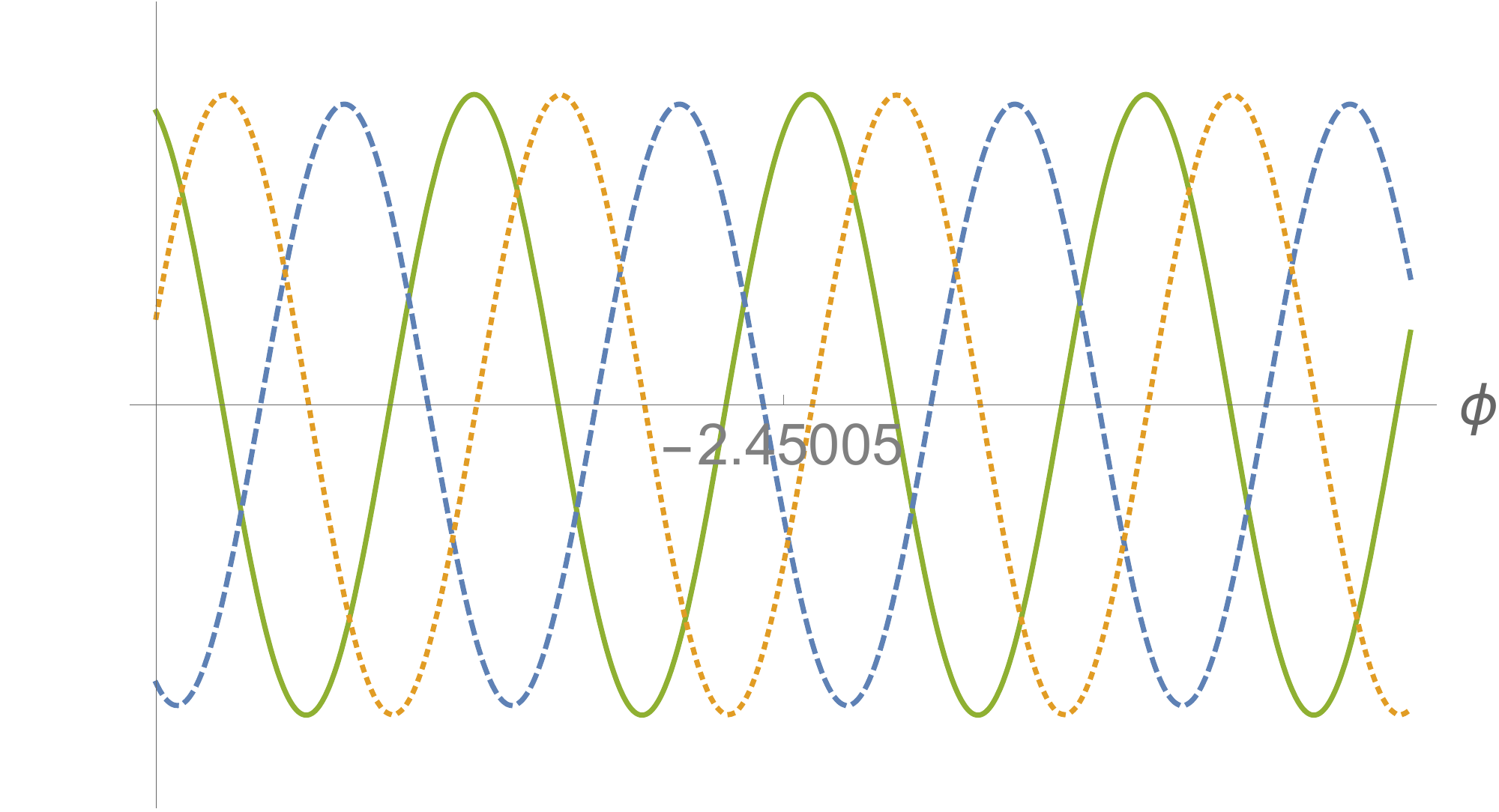}
	\includegraphics[width=0.3\textwidth]{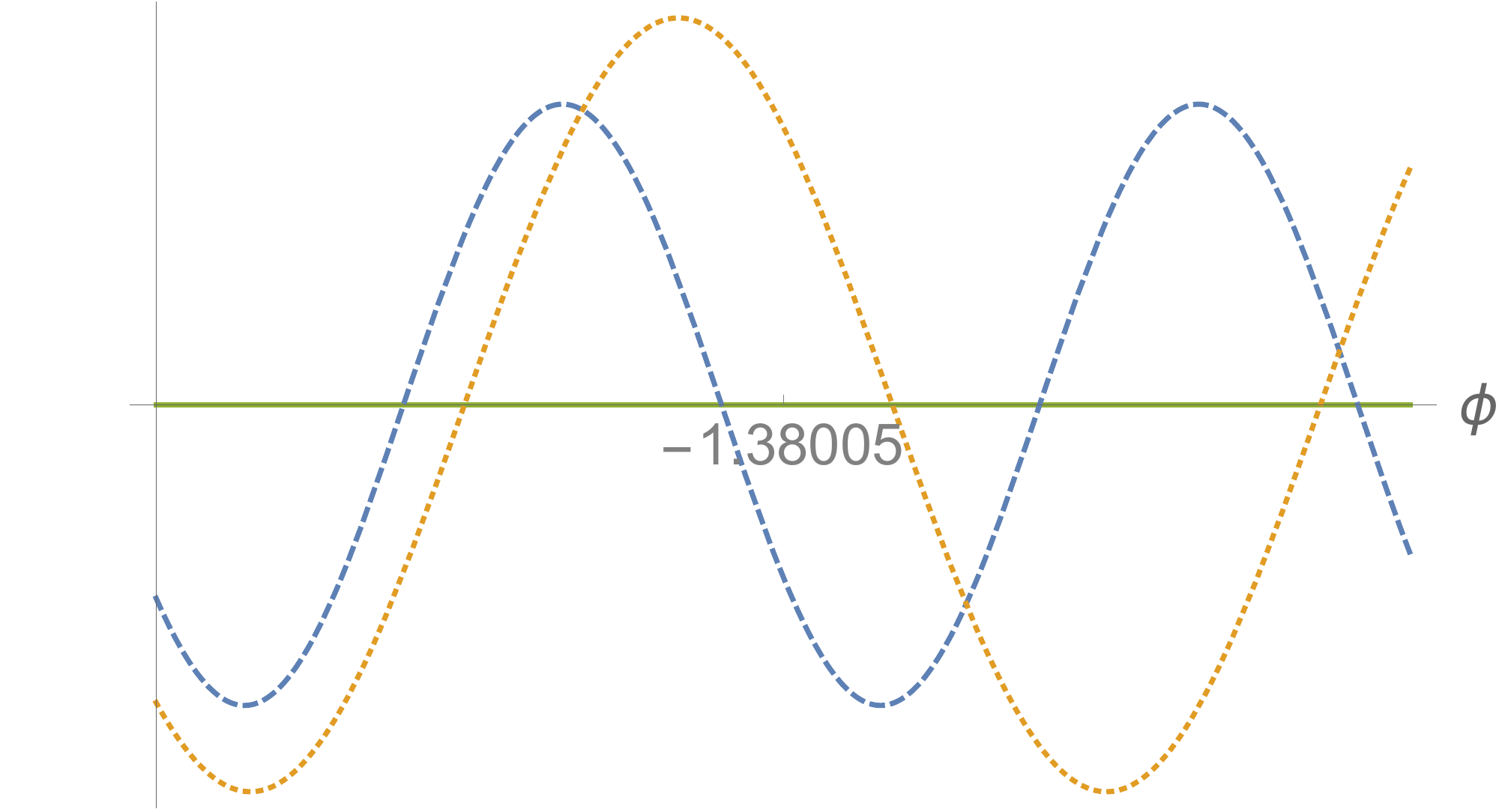}
	\caption{\label{FIG:COMP2} Comparison of the 
perturbative (blue/dashed), RO (yellow/dotted) and U.WKB wavefunctions (green/solid) for the Sauter 
pulse, with parameters $a_0=1$, $\omega=2\times 10^{-6}m$, $|p_\LCperp|=a_0 m$, $|p_z| = a_0 m/2$. 
This corresponds to $\hbar = 10^{-6}$, $\mathcal{E}=1/32$ and $\epsilon=2\times 10^{-6}$. 
\textit{Top left:} The wavefunctions agree asymptotically, far to the left of the turning 
point where $\mathcal{E}>V$. \textit{Top right:} As we approach the turning point where 
$\mathcal{E}=V$, at $\phi\simeq-1.38$, the U.WKB and RO wavefunctions agree, but the perturbative 
wavefunction begins to deviate from them. \textit{Bottom left:} Closer to the turning point, the 
RO~wavefunction also starts to deviate from the U.WKB wavefunction. \textit{Bottom right:} to the 
right of the turning point we have $\mathcal{E}<V$ and the U.WKB wavefunction is exponentially 
suppressed, whereas the perturbative and RO~wavefunctions continue to oscillate.}
\end{figure*}

Hence RO, and by extension the perturbative approach, gives an approximation of a WKB ansatz which 
is appropriate for `above the barrier' parameters. A further interpretation can be acquired by 
employing multi-scale perturbation theory~\cite{Bender:1978}. In this approach, RO captures the 
physics on only the shortest timescale and misses physics on the longer timescale associated with 
the second derivative. RO is therefore unlikely to accurately reproduce under-the-barrier physics. 
We confirm this in Fig.~\ref{FIG:COMP1} and Fig.~\ref{FIG:COMP2}, where we compare the perturbative, 
RO~and U.WKB wavefunctions for the example of Sect.~\ref{spreading}. We plot $G$ in all cases, 
obtained from $F$ by multiplying by a phase. Note that the U.WKB wavefunction is a superposition of 
perturbative or RO~wavefunctions obtained by summing contributions from $\pm p_z$ for a given~$p_z$. 
Normalisations are fixed by matching the asymptotic behaviour of the wavefunctions where the 
potential vanishes, and where they all agree. The general behaviour is the following. The three 
wavefunctions agree well when $\mathcal{E}>V$. As we approach the turning point, first the 
perturbative wavefunction and then the RO~wavefunction diverge from the U.WKB and numerical results, 
see Fig.~\ref{FIG:COMP2} for details. Beyond the turning point, under the barrier, the U.WKB 
wavefunction is exponentially damped, as it should be. Both the RO~and perturbative wavefunctions 
continue to oscillate under the barrier but with a lower frequency. The perturbative wavefunction, 
being a phase, has the same amplitude above and below, while the amplitude of the RO~wavefunction 
grows significantly under the barrier.

\begin{figure}[t!]
	\includegraphics[width=0.95\columnwidth]{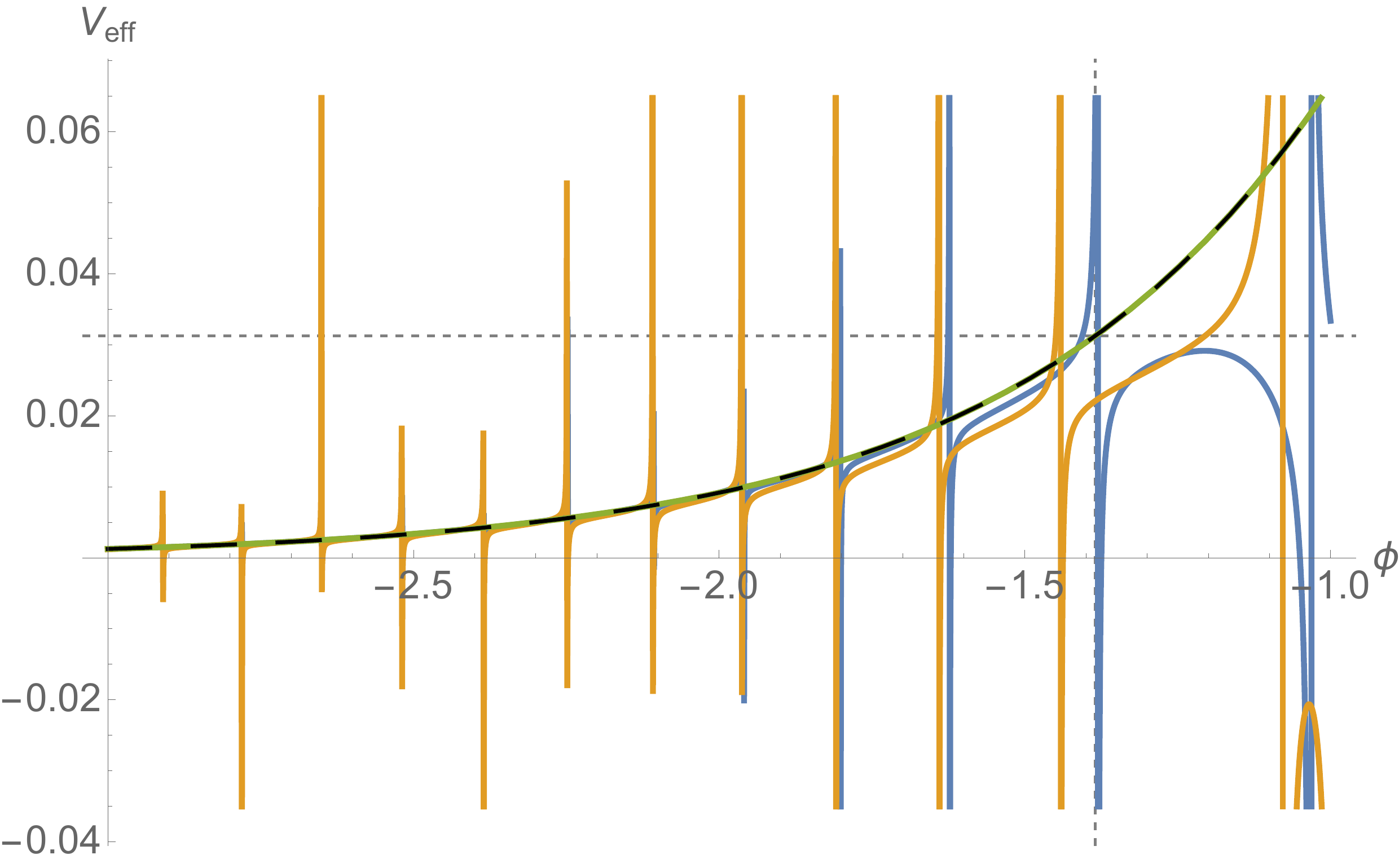}
	\includegraphics[width=0.95\columnwidth]{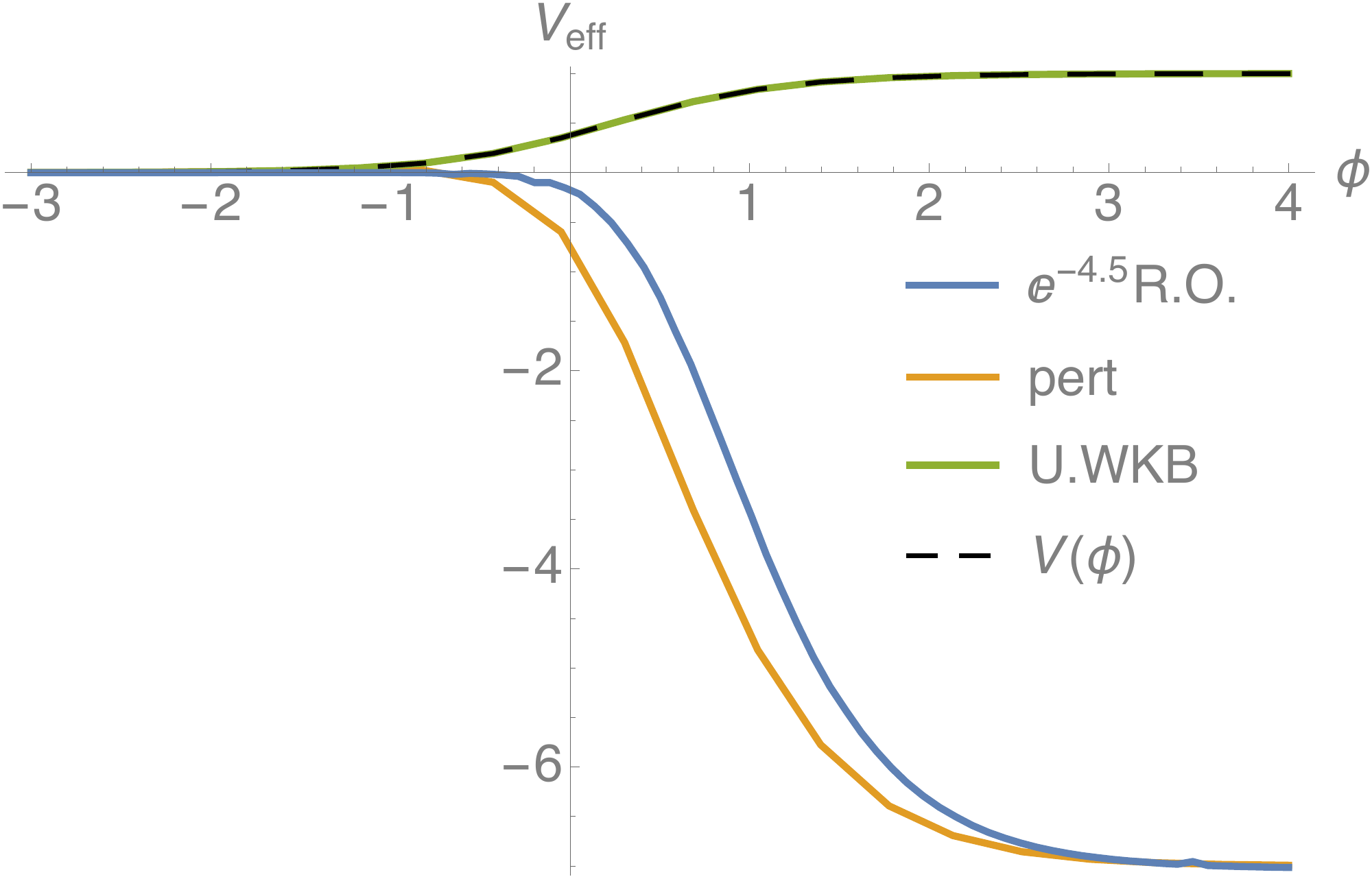}
	\caption{
	\label{FIG:VEFF}
	The effective potential (\ref{VEFF}) compared with $V(\phi)$. \textit{Top panel:} above the 
barrier, the perturbative and RO~effective potentials track $V(\phi)$, but with a series of 
divergences where $G(\phi)$ vanishes. There are no such divergences for the U.WKB potential, which 
is indistinguishable from $V(\phi)$ on the scale shown. As the wavefunctions penetrate the barrier, 
the trends change. Neglecting the divergences for clarity, the \textit{bottom panel} shows that the 
perturbative and RO~wavefunctions see a negative potential, which explains why they oscillate 
faster under the barrier than above. (The RO potential has been scaled for presentation purposes.)
	}
\end{figure}

These behaviours are unphysical. Their origin is efficiently captured by examining how well the 
Schr\"odinger equation is satisfied through consideration of an effective potential $V_\text{eff}$ 
`seen' by the wavefunction, defined by 
\be\label{VEFF}
	V_\text{eff}(\phi) := \frac{\hbar^2G''(\phi)/2+\mathcal{E}G(\phi)}{G(\phi)} \;.
\ee
For an exact solution we have $V_\text{eff}\equiv V$, from (\ref{SCHRO}). The effective potentials 
seen by the perturbative, RO~and U.WKB wavefuntions are plotted in Fig.~\ref{FIG:VEFF}~. Both the 
perturbative and RO~effective potentials become negative to the right of the turning point, and 
asymptote to a value far below, rather than above, the energy eigenvalue. This explains why the 
wavefunctions 
oscillate more quickly under the barrier: take $\phi\gg1$ so that the effective potentials are flat 
and negative, then the particles see themselves as much further above the potential than they were 
for $\phi\ll 0$, as the effective $\mathcal{E}-V(\infty) \gg \mathcal{E}$, and hence the wavenumber, 
or frequency of oscillation, increases. The particularly rapid oscillations seen in the RO~case are 
explained by the very large change in amplitude of the effective potential, due to the real 
exponential factor in (\ref{F.PERT}). This is an artefact of the approximation, in particular of 
expanding the prefactor of the WKB wavefunction.

\subsection{Barrier transmission}
Another situation in which the approximations can be tested against the well-understood physics of 
the 
Schr\"odinger equation is barrier transmission. We take the background field to be of sech form:
\be\label{sech}
	a_\mu(\phi) = a_0 m l_\mu \textrm{sech}(\phi) \equiv a_0 m 
l_\mu g(\phi)\;.
\ee
For $\kappa$ as in Sect.~\ref{spreading}, with $\kappa=-1$, we identify
\be
\frac{\hbar^{2}}{2} =  \frac{-k^2}{3a_0^2m^2} \;, \quad \mathcal{E} = 
\frac{\tilde{p}^2-m^2}{3a_0^2m^2} \; ,
\ee
and $V = (g^2+2g)/3$ has the form of a potential hill with a peak amplitude of unity.

\begin{figure}[t!!]
	\includegraphics[width=\columnwidth]{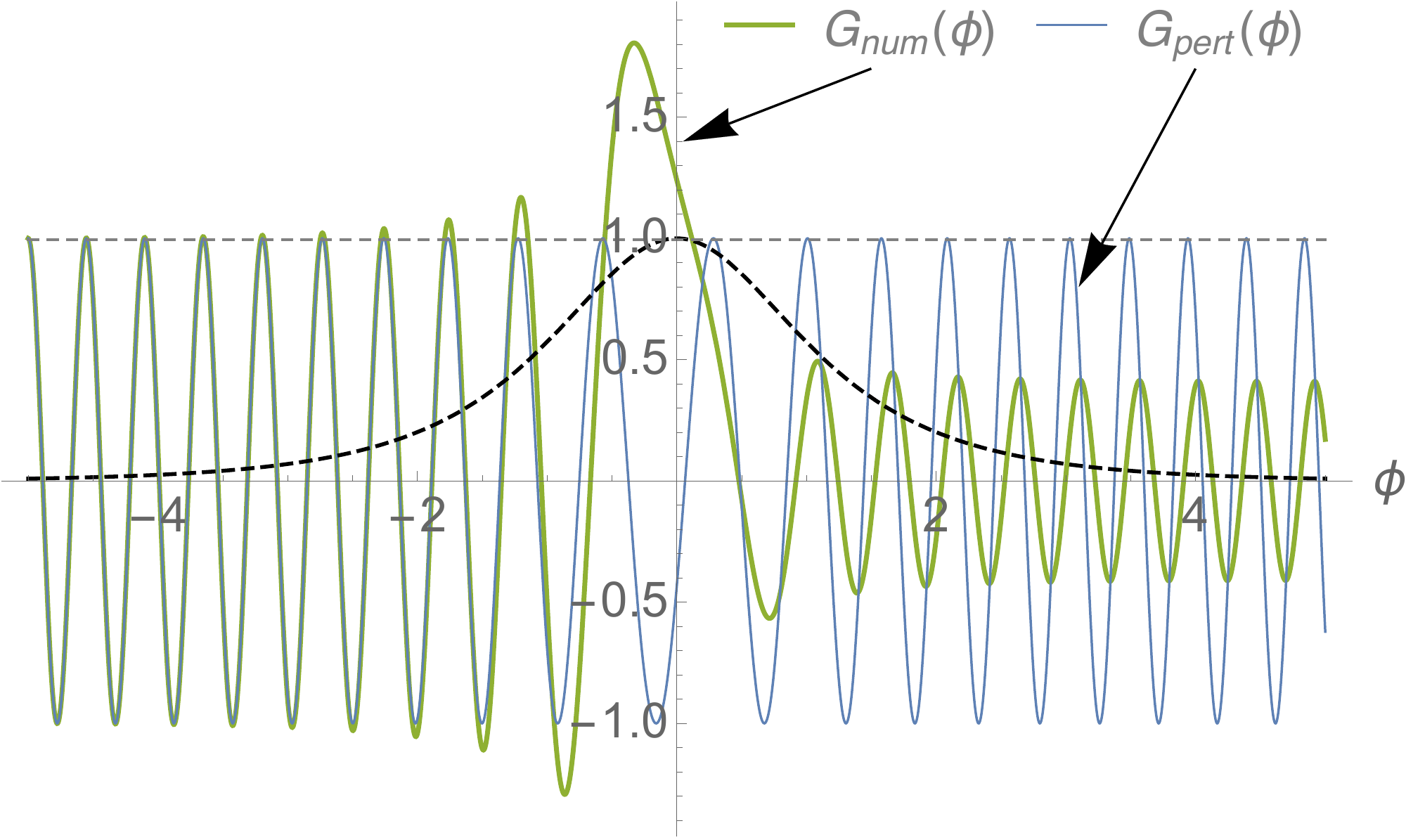}
	\includegraphics[width=\columnwidth]{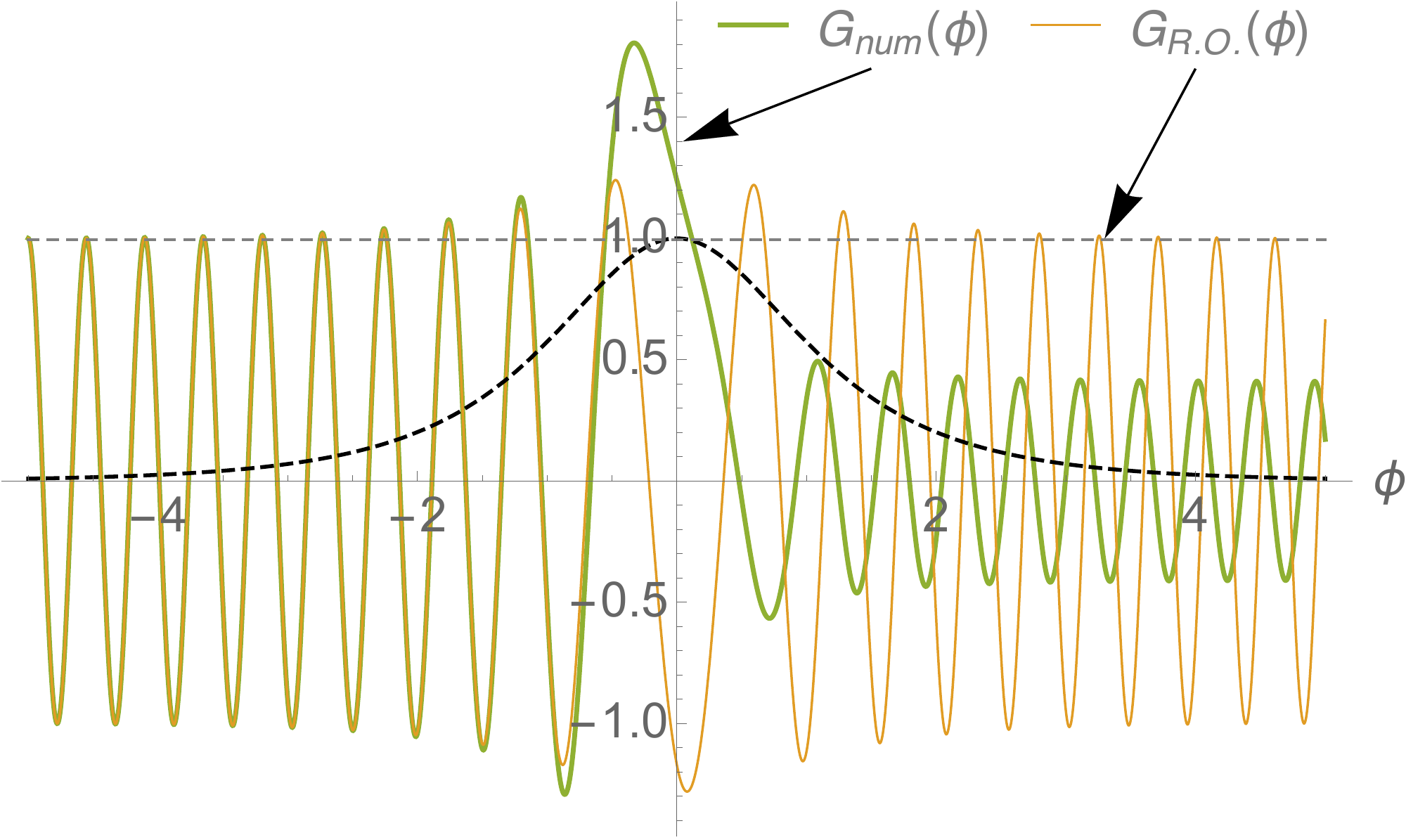}
	\caption{\label{FIG:SECH} Comparison of the numerical solution of the Schr\"odinger equation 
in the sech-type potential (\ref{sech}) with analytic approximations. $\mathcal{E}=0.995$ 
(horizontal dashed line) just below the peak of the potential (black dashed line), $\kappa=-1$, and 
initial conditions $G(-5)=1$, $G'(-5)=0$. \textit{Upper panel:}  perturbative approximation. 
\textit{Lower panel:} RO~approximation. The parameters correspond to $\epsilon\simeq 4\times 
10^{-2}$. Even when the incident energy is only just below the peak of the potential, the 
perturbative and RO~approximations fail to capture the physics of the Schr\"odinger equation.}
\end{figure}

For the parameters in Fig.~\ref{FIG:SECH}, the energy is just below the potential peak, so that 
transmission of the particle through the barrier is classically-forbidden, but only just. We 
compare 
the first-order approaches with an exact numerical solution of the Schr\"odinger equation. Initial 
conditions are chosen such that all wavefunctions agree for $\phi\ll 0$ where the potential 
vanishes. While the numerical solution shows reduced transmission as expected, the first-order 
approaches predict complete transmission through the barrier, in that the amplitudes of the 
approximate wavefunctions both to the left and right of the potential hill are equal. This suggests 
that unitarity is violated, and confirms that the first-order approximations are unable to capture 
the full physics.

\subsection{Current conservation}
%
Making an analogy with the Schr\"odinger equation has proven useful in analysing the physical 
content of the approximate wavefunctions above, but we can also consider observables directly, as in 
Sect.~\ref{SECT:MATHIEU}. Consider then the current, for a general field shape, and again in the 
case that the discriminant in (\ref{disc}) is always positive, i.e.~we are above the barrier. The 
classical current is then
\be\begin{split}\label{current}
	j_\mu(x) = -\frac{k^2}{k.p}\delta^{\underline 3}(x-x(\phi)) \frac{\pi_\mu(\phi)}{s(\phi)} 
\, 
\;,
\end{split}
\ee
where the delta function is just the product of deltas in the three directions orthogonal to $k.x$. 
We can compare this to the field theory current,
\be\label{J-DEF}
	J_\mu(x) = \frac{i}{2} \bar\varphi \overset{\leftrightarrow}{D_\mu} \varphi \;,
\ee
calculated using the various approximate wavefunctions above. Using the perturbative, RO, and WKB 
approximations gives
\be\begin{split}
	J^\text{pert}_\mu  &= p_\mu - a_\mu(\phi) + u(\phi)k_\mu \;, \\
	J^\text{RO}_\mu  &= e^{-2\epsilon u(\phi)} \big( p_\mu - a_\mu + \big[u(\phi)-\epsilon 
u^2(\phi)\big]k_\mu \big) \;, \\
	J^\text{WKB}_\mu  &= \frac{\pi_\mu(\phi)}{s(\phi)} \;. 
\end{split}
\ee
Neglecting overall normalisations, the WKB wavefunction recovers the classical current\footnote{The 
delta functions in the classical current would be recovered upon using suitably normalised 
wavepackets of solutions to the Klein-Gordon equation.}, while the RO and perturbative wavefunctions 
give only approximations to it. Higher-order WKB will add quantum corrections to the current. This 
should be contrasted with the WKB-exact plane wave case, where the classical and quantum currents 
are equal up to normalisation.

For the WKB, RO and perurbative wavefunctions, we observe that current conservation can be expressed 
as
\be\label{CC}
	\partial_\mu J^\mu = \frac{\ud}{\ud \phi} k.J = 0\;.
\ee
For over-the-barrier parameters, we note that the classical, and therefore WKB, currents are 
conserved, since
\be
	\frac{k.J^\text{WKB}}{k.p} = 1 \;.
\ee
Both the perturbative and RO~methods violate this conservation. We have
\begin{align}
	\frac{k.J^\text{pert}}{k.p} &= 1+ 2 \epsilon u(\phi) = 1 +\mathcal{O}(\epsilon u) \;, \\
	\frac{k.J^\text{RO}}{k.p} &= e^{-2\epsilon u}(1+ 2 \epsilon u - 2 \epsilon^2 u^2) = 1 
+\mathcal{O}(\epsilon^2 u^2) \;,
\end{align}
so that the current is conserved only up to a certain order in $\epsilon u = V/(4\mathcal{E}) \sim 
1/(4\mathcal{E})$, consistent with (\ref{F.OMSKRIVNING}).

\section{Conclusions}\label{SECT:CONCS}
We have examined a variety of methods of analysing the classical and quantum dynamics of charges in 
electromagnetic backgrounds depending on a single variable $k.x$, where $k_\mu$ may be lightlike, 
timelike or spacelike. The lightlike case has been the focus of great interest for over 50 years as 
it corresponds to the plane wave model of intense laser fields. The classical dynamics is then 
integrable (the Lorentz force equation is exactly soluble), as is the quantum dynamics (through the 
WKB-exact Volkov solutions of the Klein-Gordon and Dirac equations). This integrability is lost 
upon 
relaxing the null-vector condition $k^2 = 0$, however. There are many physical scenarios 
corresponding to $k^2 \ne 0$. The timelike case, $k^2 > 0$, is Lorentz equivalent to a 
time-dependent,  spatially homogeneous electric field, while the spacelike case, $k^2 < 0$, is 
equivalent to a static but inhomogeneous magnetic field.

We have first analysed classical dynamics in the spacelike case. There are three momentum 
conservation laws corresponding to translation invariance in the coordinates different from $k.x$. This, together with the mass-shell constraint, is sufficient to guarantee the existence of a first 
integral which expresses the particle momentum as a function of $k.x$. However, in contrast to the lightlike case, the appearance of a square root non-linearity means that a second integration to obtain the particle orbits is only possible in special cases.  Hence abandoning the plane wave 
nature of the background destroys integrability despite the fact that the number of (translational) 
symmetries is maintained.  We have discussed a special, solvable, case by comparing the electron radiation spectrum in a helical wiggler ($k^2 < 0$) with that in a plane wave or 
laser ($k^2 = 0$). Despite the two spectra becoming almost identical in the high energy limit, 
$\gamma \to \infty$, there is an intermediate energy range where $\gamma \gg 1$ and the spectra are significantly shifted. This is particularly relevant to simulations of electromagnetic cascades in laser backgrounds, which often adopt a (constant crossed) plane wave model at the magnetic node of a standing wave ($k^{2} \neq 0$)~\cite{Bell:2008ppp,Mironov:2009cr,Nerush:2011pp,King:2013pp}. For example, the nonlinear Compton scattering stage of a cascade becomes probable for $\gamma=50$ in an optical field with intensity parameter $a_{0} \approx 300$.

In the quantum regime, the WKB method ceases to be exact when $k^2 \ne 0$ so that there is no 
analogue of the Volkov solution. Nevertheless, approximations based on WKB and uniform-WKB 
approaches work well -- these are based on physical arguments, agree very well with numerics, and 
also produce the expected physics. The optimal uniform-WKB ansatz depends strongly both on field 
shape and parameters. 

We have also considered perturbation theory in kinematically small parameters, leading to the use of 
`reduction of order', familiar from the study of radiation reaction. This is a general approach in 
that it does not make reference to the shape of the background field considered. This seems like a 
promising method of generalising the Volkov solution, at least approximately, but we have seen that 
it is essentially a high energy approximation~\cite{Blankenbecler:1987rg} which is limited to `above 
the barrier' type problems. This is particularly relevant in the context of laser-matter 
interactions, where reduction of order is implicitly invoked whenever the Lorentz-Abraham-Dirac 
equation is replaced by that of Landau and Lifshitz, and suggests that reduction of order should be 
further investigated in order to establish its range of validity.

\acknowledgements
A.I.~thanks Gerald Dunne for useful discussions, and the Centre for Mathematical Sciences, Plymouth 
University, for hospitality. A.I.~is supported by the Olle Engkvist Foundation, contract 2014/744, 
and the Lars Hierta Memorial Foundation, grant FO2015-0359.

\end{document}